\documentstyle[harvard,epsfig]{mn}

\input{epsf}

\def\lsim{\mathrel{\hbox{\rlap{\hbox{\lower4pt\hbox{$\sim$}}}\hbox{$<$}}}}
\def\gsim{\mathrel{\hbox{\rlap{\hbox{\lower4pt\hbox{$\sim$}}}\hbox{$>$}}}}

\def\and   {\rm {et al.} \rm}

\begin{document}

\title
[Constraining Cosmology from Redshift Space Distortions via {\bf $\xi(\sigma,\pi)$}]
{
The 2dF QSO Redshift Survey - VII. Constraining Cosmology from Redshift Space Distortions via  $\xi(\sigma,\pi)$ }

\author[F. Hoyle  et al ]
{
Fiona Hoyle$^{1,2,*}$, P.~J. Outram$^1$, T. Shanks$^1$, B.~J. Boyle$^{3}$, S. M. Croom$^{3}$, 
\newauthor \&  R.~J. Smith$^{4}$
\\
1 Department of Physics, Science Laboratories, South Road, Durham, DH1 3LE, U.K \\
2 Department of Physics, Drexel University, 3141 Chestnut Street, Philadelphia, PA 19104 \\
3 Anglo-Australian Observatory, PO Box 296, Epping, NSW 2121, Australia\\
4 Liverpool John Moores University, Twelve Quays House, Egerton Wharf, Birkenhead, CH41 1LD \\
* email hoyle@venus.physics.drexel.edu\\
}

\maketitle 

\begin{abstract}

We describe a method from which cosmology may be constrained from the 2QZ Survey. By comparing clustering properties parallel and perpendicular to the line of sight and by modeling the effects of redshift space distortions, we are able to study geometric distortions in the clustering pattern which occur if a wrong cosmology is assumed when translating redshifts into comoving distances. Alcock and Paczy\`{n}ski pointed out that this technique was particularly sensitive as a test for a large cosmological constant, $\Lambda$.

Using mock 2QZ catalogues, drawn from the {\it Hubble Volume} simulation, we 
find, that there is a degeneracy between
the geometric distortions and the redshift-space distortions, parameterised by 
$\beta_{QSO}(\bar{z})$, that makes it difficult to obtain an unambiguous 
estimate of  $\Omega_{\rm m}(0)$, the matter density parameter, from the 
geometric tests alone. This is in agreement with the conclusions of 
Ballinger et al. However,
we demonstrate a new method to determine the cosmology which works by
combining the above geometric test with a test based on the evolution of
the QSO clustering amplitude, which has a different dependence on
$\beta_{QSO}(\bar{z})$ and $\Omega_{\rm m}(0)$. 
In the analysis of the Hubble Volume mock 
catalogues we find that we are able to break the degeneracy between 
$\Omega_{\rm m}(0)$ and $\beta_{QSO}(\bar{z})$ and that independent 
constraints to $\pm20$\% (1$\sigma$)
accuracy on $\Omega_{\rm m}(0)$ and $\pm10$\% (1$\sigma$) accuracy 
on $\beta_{QSO}(\bar{z})$ should be 
possible in the full 2QZ survey.

Finally we apply the method to the {\it 10k catalogue} of 2QZ QSOs. The smaller number of QSOs and the current status of the Survey mean that a strong result on cosmology is not possible but we do constrain $\beta_{QSO}(\bar{z})$ to $0.35\pm0.2$. By combining this constraint with the further constraint available from the amplitude of QSO clustering, we find tentative evidence favouring a model with non-zero $\Omega_{\Lambda}(0)$, although an $\Omega_{\rm m}(0)=1$ model provides only a marginally less good fit. A model with $\Omega_{\Lambda}(0)=1$ is ruled out. The results are in agreement with those found by Outram et al. using a similar analysis in Fourier space.

\end{abstract}


\section{Introduction}

One of the challenges that still faces cosmologists is to determine the values of the underlying cosmological parameters of the Universe, such as the density parameter, $\Omega_{\rm m}(0)$, and the energy density associated with the cosmological constant, $\Omega_{\Lambda}(0)$. During the 1990's, it became clear that there was a problem with the favoured Einstein-de Sitter, $\Omega_{\rm m}(0)$=1 model. Hubble's constant was being measured as $H_0 \sim $ 70 km $\; s^{-1}$Mpc$^{-1}$ which gives an age for the Universe of $\sim 10$ Gyr if $\Omega_{\rm m}(0)$=1. Globular clusters were estimated to be older than the Universe itself at 16$\pm$2 Gyr \cite{Renz96}, although recent estimates of the ages of Globular clusters are slightly lower, e.g. 11.8$\pm$2.1 Gyr \cite{Grat97}. However, inflationary theory, in its simplest form, implies a flat universe. To comply with this requirement, a cosmology where $\Omega_{\rm m}(0) + \Omega_{\Lambda}(0)$=1 was suggested as an alternative to the Einstein-de Sitter model \cite{Peebles84}. Invoking a cosmological constant, $\Lambda$ term ($\Omega_{\Lambda}(0) = \Lambda c^2/(3H_0^2)$), provides a solution to the problem of the age of the Universe, since with lower values of $\Omega_{\rm m}(0)$ (and $\Omega_{\rm m}(0) + \Omega_{\Lambda}(0)$=1), the age of the Universe is increased. A cosmology with non-zero $\Lambda$ also matches the amplitude of fluctuations on both COBE \cite{smoot92} and cluster scales \cite{Eke96}. However, the physics underpinning  such a small $\Lambda$ term remains very unclear (e.g. Witten 2000).

One of the most elegant ways of testing for the existence of a non-zero $\Lambda$ term was put forward by \citeasnoun{AP79}. They suggested that if the clustering of galaxies or QSOs, parallel and perpendicular to the line of sight, was computed assuming an Einstein-de Sitter cosmology but the Universe had a different cosmology, for example one which had $\Omega_{\Lambda}(0) \neq 0$, then a squashing or elongation of the clustering pattern would occur more in one direction than in the other. This is because assuming the wrong cosmology affects the clustering pattern parallel to the line of sight differently to that perpendicular to the line of sight. The cosmology {\it assumed} for measurements of the clustering can have any value of $\Omega_{\rm m}(0)$ and $\Omega_{\Lambda}(0)$ but if these values differ from the {\it underlying cosmology} of the Universe, distortions are introduced into the clustering pattern. The only assumption required is that the clustering in {\it real space} is on average spherically symmetric.

Clustering statistics measured from galaxy or QSO redshift surveys are, however, not measured in real space and so they have redshift space distortions imprinted on them. On small scales, virialized clusters appear elongated along the line of sight in redshift space. These are the so called `fingers of God'. They arise because the redshift due to cosmological expansion and redshift due to peculiar motions cannot be separated. The redshift is converted into a distance assuming that there are no peculiar motions, which distorts the shape of the cluster. On large scales, coherent infall squashes over-densities along the line of sight in redshift space. This causes a boost in the amplitude of the redshift space correlation function on linear scales, as compared to the real space correlation function. This is characterised by the parameter $\beta = \Omega^{0.6}_{\rm m}/b$ \cite{Kaiser87}. In order to measure properly any effect of $\Lambda$, these two contributions have to be accurately accounted for in any model of the clustering. Models for the redshift space distortions have appeared many times in the literature \citeaffixed{Bean83,BPH96,MS96,Pop98,Rat98c}{e.g.}, and we describe some of these models in section \ref{sec:model}.

Either galaxy or QSO surveys can, in theory, be used to detect the distortion introduced into the clustering pattern through incorrect assumptions for the cosmology. However, wide angle galaxy surveys, including the 2dF Galaxy Redshift Survey, essentially probe the Universe at $z=0$, where assuming the incorrect cosmology would make little difference to the measured clustering pattern. Clustering from pencil beam surveys, such as the CNOC survey \cite{Yee00}, cannot be measured over a wide range of scales. QSO surveys sample space more sparsely than current galaxy surveys but probe clustering out to high redshifts over a wide area, potentially allowing the distortions in the clustering to be detected.

The aims of this paper are two fold. First we use the mock catalogues to test how well $\Omega_{\rm m}(0)$, $\Omega_{\Lambda}(0)$ and $\beta_{QSO}(\bar{z})$ will be constrained from the 2dF QSO Redshift Survey (2QZ). At this stage, we only consider flat cosmologies with $\Omega_{\rm m}(0) + \Omega_{\Lambda}(0)$=1, as expected from current CMB experiments \cite{balbi00,debern00} and inflationary theory. The method can, however, be extended to allow for open or closed cosmologies. We also test whether tighter constraints on the cosmology can be found if the results from $\xi(\sigma,\pi)$ are combined with results from the evolution of the QSO-mass bias. Second, we apply the method to the 2QZ {\it 10K Catalogue} \cite{Scott10k} to obtain preliminary results on cosmology and demonstrate the feasibility of this method when applied to real data.

The outline of this paper is as follows. In section \ref{sec:data}, we describe both the 2QZ Survey and the {\it Hubble Volume} simulation, from which the mock catalogues are constructed. In section \ref{sec:meas}, we discuss how the correlation function parallel and perpendicular to the line of sight, $\xi(\sigma,\pi)$, is measured from either the simulation or the 2QZ Survey and in section \ref{sec:model} we discuss how we model $\xi(\sigma,\pi)$. In section \ref{sec:true}, we outline our method for detecting cosmology and in section \ref{sec:reslts} we present our results. In section \ref{sec:evo}, we consider whether other constraints can be usefully combined with the results from $\xi(\sigma,\pi)$. In Section \ref{sec:10k} we apply the analysis to the {\it 10k catalogue} \cite{Scott10k} and finally in section \ref{sec:cnc} we draw our conclusions.

\section{2QZ Survey and Mock Catalogues} 
\label{sec:data}

\subsection{2QZ}

The 2QZ redshift survey aims to measure the redshifts of 25,000 QSOs over a redshift range of 0.3$\lsim z \lsim$ 3. The spectra of the QSOs will be measured using the 2dF instrument on the Anglo-Australian Telescope (AAT). This instrument allows up to 400 spectra to be obtained simultaneously, allowing a large number of QSOs to be observed in a relatively short period of time. QSOs are selected in a homogeneous manner via {\it ub$_{\rm J}$r} multi-colour selection and the survey will be more than 90\% complete to $z \sim$2.2 \cite{Boyle00}. The spectra of the objects are obtained using the 2dF 
instrument on the AAT. The data is reduced using the 2dF pipeline reduction
system \cite{Karl} and objects are
identified as QSOs by an automated procedure known as {\tt AUTOZ} 
\cite{Miller00}. {\tt AUTOZ} also identifies the QSO redshifts; these 
are then visually confirmed by two independent observers.

Two strips of the sky are being observed, each measuring 75$^{\circ} \times 5^{\circ}$. One is centred at $\delta = -30^{\circ}$, with 21$^{\rm h}$40$^{\rm m}$ $\lsim \alpha \lsim $ to 03$^{\rm h}$15$^{\rm m}$, close to the Southern Galactic Pole, while the other is centred at $\delta= 0^{\circ}$ with 09$^{\rm h}$50$^{\rm m}$ $\lsim \alpha \lsim $ to 14$^{\rm h}$50$^{\rm m}$ in the North Galactic Cap. The total area will be around 740 square degrees.

For more details see \citeasnoun{Croomconf}, \citeasnoun{Boyle00}, \citeasnoun{Scott00}, \citeasnoun{HoyleQSOPK}, \citeasnoun{Scott10k}, \citeasnoun{Smith01} and also {\tt www.2dfquasar.org}.

\subsection{The Hubble Volume Simulations}

The {\it Hubble Volume} Simulations \footnote{
The Hubble Volume simulations were performed
by the ``Virgo consortium for cosmological simulations''.
This is an international collaboration involving universities
in the UK, Germany and Canada.
The members of this consortium are: J. Colberg, 
H. Couchman, G. Efstathiou, C. Frenk (PI), A. Jenkins, A. Nelson,
J. Peacock, F. Pearce, P. Thomas, and S. White. G. Evrard is an 
associate member. The Hubble Volume simulation was carried out on the
Cray-T3E at the Max-Planck Rechen Zentrum in Garching.}
 are a series of simulations run by the Virgo Consortium and to date are the largest N-body simulations ever (see \citeasnoun{Jenkins98} and \citeasnoun{Evrard} for more details on Virgo consortium simulations). In order to make realistic mock catalogues, a simulation where the dark matter particles are output along an observer's past lightcone is required. The $\Lambda$CDM simulation has one deep lightcone output which covers an area of 15 $\times$ 75 degrees and extends out to $z \sim$ 4.

The parameters of the simulation are $\Omega_{\rm m}(0)$=0.3, $\Omega_{\Lambda}(0)$=0.7, $H_{\circ}$=70 km s$^{-1}$Mpc$^{-1}$ and the normalisation, $\sigma_8$, is 0.9, consistent with the abundance of hot X-ray clusters \cite{WEF93} and with the level of anisotropies in the CMBR found by COBE \cite{smoot92}. The input power spectrum was calculated using CMBFAST \cite{CMBFAST} with the parameters listed above but assuming that $\Omega_{\rm b(0)}$=0.04 and $\Omega_{\rm CDM(0)}$=0.26. This changes the shape of the input power spectrum to $\Gamma = \Omega_{\rm m}(0) h \;{\rm exp}( -\Omega_{\rm b(0)} [\sqrt{2h} + \Omega_{\rm m}(0)]/\Omega_{\rm m}(0))$=0.17 \cite{Sugi95} as compared to $\Gamma=\Omega_{\rm m}(0) h = 0.21$.

One billion mass particles are contained within a cube that is 3,000$h^{-1}$Mpc on a side. One of the vertices was chosen to be the observer and the long axis of the lightcone was oriented along the maximal diagonal. The lightcone therefore extends to a depth of $\sim$5,000$h^{-1}$Mpc, which corresponds to $z \sim $4 in the $\Lambda$CDM cosmology. The solid angle of the lightcone is 75$\times$15 degrees which is split into three 75$\times$5 degree slices. The 2QZ Survey consists of 2 such slices. Ideally we would like many more than 3 slices but due to the large volume of the survey (4.2$\times 10^{9} h^{-3}$Mpc$^3$ for $z < 2.2$ and $\Omega_{\rm m}(0)$=0.3, $\Omega_{\Lambda}(0)$=0.7) this is currently not possible.

\subsection{Construction of the Mock Catalogues}

To create realistic mock catalogues, there are a number of steps which must be carried out. First the mass particles must be biased in order to match the predicted clustering amplitude of 2QZ QSOs. Initial analysis of the correlation function, assuming a cosmology with $\Omega_{\rm m}(0)=0.3$, $\Omega_{\Lambda}(0)=0.7$, found that the correlation function could be well approximated by a power law of the form $\xi(s) = (s/s_0)^{-\gamma}$, with $\gamma \approx 1.7$ and $s_0 \approx 6.0 h^{-1}$Mpc, at all epochs, consistent with the work of \citeasnoun{croomshanks}. See \citeasnoun{Scott00} for the correlation function of the 10k catalogue of 2QZ QSOs.

The biasing prescription used in this paper is similar to the method described in \citeasnoun{Hoyle99} and follows method 2 as described by \citeasnoun{Cole98}. The bias probability is based on the density field at the epoch at which particles are selected rather than from the initial density field. The bias is approximately scale independent on the scales we probe here. A full description of the biasing procedure can be found in \citeasnoun{PhD} and \citeasnoun{HoyleQSOPK}. 

We also have to match the radial selection function of the QSOs. We fit a polynomial to the $N(z)$ distribution of the QSOs with redshifts in the range $0.3 < z < 2.2$ to obtain a probability distribution function and randomly select particles using this until there are $\approx$12,500 mock QSOs on each slice.

To mimic the two independent slices of the 2QZ, we measure $\xi(\sigma,\pi)$ from the two outer slices and average the results together.

\section{Measuring {\bf $\xi(\sigma,\pi)$}}
\label{sec:meas}

There are many different ways in which the clustering perpendicular ($\sigma$) and parallel ($\pi$) to the line of sight can be defined. Here, we define $\pi = |s_2 - s_1|$ and $\sigma = (s_1 + s_2)\theta/2$, where $s_1$ and $s_2$\footnote{We adopt the convention that $s$ refers to apparent distances in redshift space and $r$ refers to distances in real space through out this paper.}  are the distances to two QSOs and $\theta$ is the angle between them, measured from the position of an observer. 

$\xi(\sigma,\pi)$ is then estimated in much the same way as the two-point correlation function. A catalogue of unclustered points, that have the same radial selection function and angular mask as the data but which contains many more points than the data catalogue, is used to estimate the effective volume of each bin. The $DD(\sigma,\pi)$, $DR(\sigma,\pi)$ and $RR(\sigma,\pi)$ (where $D$ stands for data or mock QSO and $R$ stands for random) counts in each $\pi$ and $\sigma$ bin are found and the Hamilton estimator \cite{Ham93} 
\begin{equation}
\xi(\sigma,\pi) = \frac{DD(\sigma,\pi)RR(\sigma,\pi)}{DR(\sigma,\pi)^2} - 1
\end{equation}
is used to find $\xi(\sigma,\pi)$. Due to the sparsity of QSOs in the 2QZ Survey, we use bins of $\delta$log($\pi/h^{-1}$Mpc) = $\delta$log($\sigma/h^{-1}$Mpc) = 0.2.  The sparsity of QSOs also means that Poisson errors are a reasonable estimate of the error on scales less than $40 h^{-1}$Mpc \cite{PhD}.
\begin{equation}
\Delta \xi(\sigma,\pi) = (1 + \xi(\sigma,\pi)) {\sqrt \frac{2}{DD(\sigma,\pi)}}.
\label{eq:pois}
\end{equation}
The factor of two is needed as we require the number of independent pairs in each bin. 

\section{Modeling {\bf $\xi(\sigma,\pi)$}}
\label{sec:model}

\subsection{Definitions}
\label{sec:defs}

There are many terms used in this analysis that can be easily confused. We define their meaning here and stick to these conventions throughout.

\begin{itemize}
\item {Underlying cosmology - this is the (unknown) cosmology of the Universe}

\item{The simulation cosmology - the known cosmology ($\Omega_{\rm m}(0)$=0.3, $\Omega_{\Lambda}(0)$=0.7) of the {\it Hubble Volume} Simulation}

\item {Assumed cosmology - the cosmology used when measuring the two-point correlation function and $\xi(\sigma,\pi)$ from the 2QZ Survey or the {\it Hubble Volume} simulation. Models of $\xi(\sigma,\pi)$, which are discussed later, also have to be calculated in the same assumed cosmology as the data. We consider two possibilities for the assumed cosmology, $\Omega_{\rm m}(0)$=1, $\Omega_{\Lambda}(0)$=0 (EdS) or $\Omega_{\rm m}(0)$=0.3, $\Omega_{\Lambda}(0)$=0.7 (the $\Lambda$ cosmology), when fitting the models to the simulation to show that the results are not sensitive to this choice.}

\item{Test cosmology - the cosmology used to generate the model predictions for $\xi(\sigma,\pi)$ which are then translated into the assumed cosmology. When the test cosmology matches the underlying (or simulation) cosmology, the distortions introduced into the clustering pattern should be the same in the model and in the data. The model should then provide a good fit to the data, providing the redshift space distortions have been properly accounted for. If the test cosmology is incorrect then the model should not fit the data. This potentially allows the underlying cosmology to be determined.}
\end{itemize}

\subsection{Requirements of the Model}

One of the aims of this paper is to test whether the cosmological parameters, $\Omega_{\rm m}(0)$, $\Omega_{\Lambda}(0)$ and also $\beta_{QSO}(\bar{z})$ will be constrained from the 2QZ Survey. $\beta_{\rm g}(0)$ has been measured from galaxy redshift surveys by comparing the zeroth and second order moments of the correlation function \cite{Rat98c}. However, the effects of cosmology add an extra distortion to the clustering pattern of objects with a wide range of redshifts. This alters the moments of the clustering too, such that the effects of redshift space and cosmological distortions are difficult to disentangle. 

Instead, we test if cosmology and $\beta_{QSO}(\bar{z})$ will be constrained by comparing $\xi(\sigma,\pi)$ measured from the mock catalogues to models of $\xi(\sigma,\pi)$. The actual procedure is described in more detail in section \ref{sec:true}. The idea is that when the values of $\Omega_{\rm m}(0)$, $\Omega_{\Lambda}(0)$ and $\beta_{QSO}(\bar{z})$ used to calculate the models are the same as the underlying values, the model will match the mock catalogues, allowing the cosmology to be determined.

There are several assumptions that go into the models. One of the assumptions is that we are comparing the models to the data on linear scales so the effects of non-linear clustering can be ignored in the models. A second assumption we make is that the bias factor, $b$, is independent of scale. This is the case, by design, in the mock catalogues but this may only be an approximation on small scales in the 2QZ Survey. 

The literature contains many examples of models for $\xi(\sigma,\pi)$  \citeaffixed{MS96,Pop98,Rat98c}{e.g.} and the power spectrum, measured parallel and perpendicular to the line of sight, $P(k_{\parallel}, k_{\perp})$, \cite{BPH96}. We consider various aspects of these models as the models must account for the effects of cosmology and the redshift space distortions. We describe the method for including the cosmology below and describe a model for the effects of redshift space distortions in section \ref{sec:arat}.

\subsection{Effect of {\bf $\Lambda$} on Clustering Anisotropy}
\label{sec:lambda}

If the assumed cosmology is different from the underlying cosmology of the Universe (or the simulation), distortions, different from those caused by peculiar velocities, are introduced into the clustering pattern. This is because the radial and perpendicular directions are affected by cosmology in different ways. \citeasnoun{BPH96} outline how the cosmology scales the power spectrum split into its parallel and perpendicular components. Cosmology similarly affects the correlation function in the linear regime according to:
\begin{equation}
\xi_{\rm assumed}(\sigma,\pi) = \xi_{\rm true}(\sigma^{\prime},\pi^{\prime}),
\end{equation}
i.e. the correlation function in the assumed cosmology is the same as the correlation function in the test cosmology provided $\sigma$ and $\pi$ are scaled according to
\begin{equation}
\sigma^{\prime} = \frac{\sigma}{f_{\perp}} = \frac{\sigma B_{\rm a}}{B_{\rm t}}
\label{eq:sigscale}
\end{equation}
and
\begin{equation}
\pi^{\prime} = \frac{\pi}{f_{\parallel}} = \frac{\pi A_{\rm a}}{A_{\rm t}},
\label{eq:piscale}
\end{equation}
where the subscripts a and t refer to the assumed cosmology and the test cosmology respectively. For a flat Universe, $A$ and $B$ are defined as 
\begin{equation}
A = \frac{c}{H_0} \frac{1}{\sqrt{\Omega_{\Lambda}(0) + \Omega_{\rm m}(0)(1 + z)^3}} 
\end{equation}
and 
\begin{equation}
B = \frac{c}{H_0} \int_0^z \frac{dz^{\prime}} {\sqrt{\Omega_{\Lambda}(0) + \Omega_{\rm m}(0)(1 + z^{\prime})^3}}.
\end{equation}
These can be calculated for open universes too --- see \citeasnoun{BPH96} for the full definition --- however we only consider flat universes. Both $A$ and $B$ are calculated at the median redshift of the survey, which in this case is $z=1.4$. By comparing the models to the simulation, we find that this is an adequate approximation to make (see Figure \ref{fig:HVsigpi}).

\subsection{Redshift Space Distortion Model}
\label{sec:arat}

Following \citeasnoun{Peebles80} and \citeasnoun{Rat98c} we define the correlation function, parallel and perpendicular to the line of sight by
\begin{equation}
1 + \xi(\sigma,\pi) = \int  [1 + \xi(r)]g({\bf r,w}) {\rm d}w^3 ,
\label{eq:start}
\end{equation} 
where $\xi(r)$ is the real space QSO or mock QSO correlation function, free from the effects of redshift space distortions. ${\bf w} = {\bf v_i - v_j}$ where ${\bf v}$ is the peculiar velocity of a QSO after subtracting off the Hubble flow and ${\bf r = r_i - r_j}$. $g({\bf r,w})$ is the distribution function of ${\bf w}$ for QSO's separated by ${\bf r}$. Here $r^2 = \sigma^2 + r_z^2$, where $r_z = \pi - w_z /H$ and $w_z$ is the component of ${\bf w}$ parallel to the line of sight, denoted by $z$ for simplicity.

If it can be assumed that $g({\bf r,w})$ is a slowly varying function of ${\bf r}$ then $g({\bf r,w})$ = $g({\bf w})$. Then equation \ref{eq:start} can be simplified and becomes
\begin{equation}
1 + \xi(\sigma,\pi) = \int _{- \infty} ^{\infty} [1 + \xi(r)]f(w_z) {\rm d}w_z .
\end{equation}
where
\begin{equation}
f(w_z) = \int {\rm d}w_x \int {\rm d}w_y g({\bf w}) 
\label{eq:approx}
\end{equation}

A simple streaming model for the bulk motions of galaxies can be included by writing $g({\bf r,w}) = g[{\bf w - \hat{r}}v(r)]$ where $v(r)$ is the mean relative radial velocity of galaxies separated by ${\bf r}$. If the approximation for $g({\bf r,w})$ in equation \ref{eq:approx} is made again, equation \ref{eq:start} becomes
\begin{equation}
1 + \xi(\sigma,\pi) = \int _{- \infty} ^{\infty} [1 + \xi(r)] f[w_z - v(r_z)] dw_z .
\label{eq:sigpi}
\end{equation}

Models for the effects of small scale peculiar velocities and the bulk motions of galaxies or QSOs are required if $\xi(\sigma,\pi)$ is to be accurately described.

\subsubsection{Fingers of God}
The distribution function of the small scale peculiar velocities has been previously modeled as an exponential or a Gaussian distribution. \citeasnoun{Rat98c} found that a substantially better fit to peculiar velocities in $N$-body simulations was found if an exponential model was used and that is what we adopt here:

\begin{equation}
f(w_z) = \frac{1}{\sqrt{2} <w_z^2>^{1/2}} {\rm exp} \left \{  - \sqrt{2} \frac{|w_z|}{<w_z^2>^{1/2}} \right \},
\end{equation}
where $<w_z^2>^{1/2}$ is the rms line of sight pairwise velocity dispersion. 

Errors in the measurements of the QSO redshifts may also affect the measurement of $\xi(\sigma,\pi)$ on small scales. These errors may lead to the peculiar velocities being better modeled by a Gaussian distribution. We discuss this further in Section \ref{sec:10kHVMC}.

\subsubsection{Bulk Flows}
The model that we use for the bulk motions depends on the cosmology of the Universe, the clustering of QSOs and the QSO bias. Following \citeasnoun{DHS} we set
\begin{equation}
v(r_z) = -\frac{2}{3 - \gamma} \Omega_{\rm m}(0)^{0.6}H(z) r_z \left\{ \frac{\xi(r)}{b^2 + \xi(r)} \right\}.
\label{eq:bulk}
\end{equation}
$r_z$ is defined above, $\beta$ and $\Omega_{\rm m}(0)$ are free parameters and the bias, $b$, is calculated using $b = \Omega_{\rm m}(0)^{0.6}/\beta$ \cite{Kaiser87}. 
Clearly in the case of the QSOs, b is the QSO bias and b, $\Omega_m$ 
and $\beta$ are defined at the average QSO redshift.
The value of $H_{\circ}$ in the simulation is 70 km s$^{-1}$ Mpc$^{-1}$, which we will assume when we analyse the data too, and the evolution of Hubble's constant with redshift and flat cosmology is given by
\begin{equation}
H(z) = H_0 \left\{ \Omega_{\rm m}(0)(1 + z)^3 + \Omega_{\Lambda}(0) \right\}^{0.5}. 
\end{equation}
We self-consistently determine the real space correlation function from the redshift space correlation function. This is discussed in section \ref{sec:true}. The real space correlation function is not approximated by a power law but a value of $\gamma$ is still required in equation \ref{eq:bulk}. A value of $\gamma$=1.7 best fits the redshift space correlation function over the range $4 < s < 40 h^{-1}$Mpc \cite{PhD} so this is the value used in equation \ref{eq:bulk}.

An alternative method for modeling the effects of infall on $\xi(\sigma,\pi)$ is outlined in \citeasnoun{MS96}. Essentially, they generalise the formula of \citeasnoun{Ham92} to define a relation between the redshift space correlation function at $z\neq0$ and the real space correlation function. This model includes the effects of infall but not small scale peculiar velocities. These could be included by convolving the redshift space correlation function with an exponential model for small scale peculiar velocities along the $\pi$ direction. 

\subsection{Comparison with the Hubble Volume Simulation}

\begin{figure*} 
\begin{centering}
\begin{tabular}{ccc}
{\epsfxsize=5.7truecm \epsfysize=5.truecm \epsfbox[10 10 550 560]{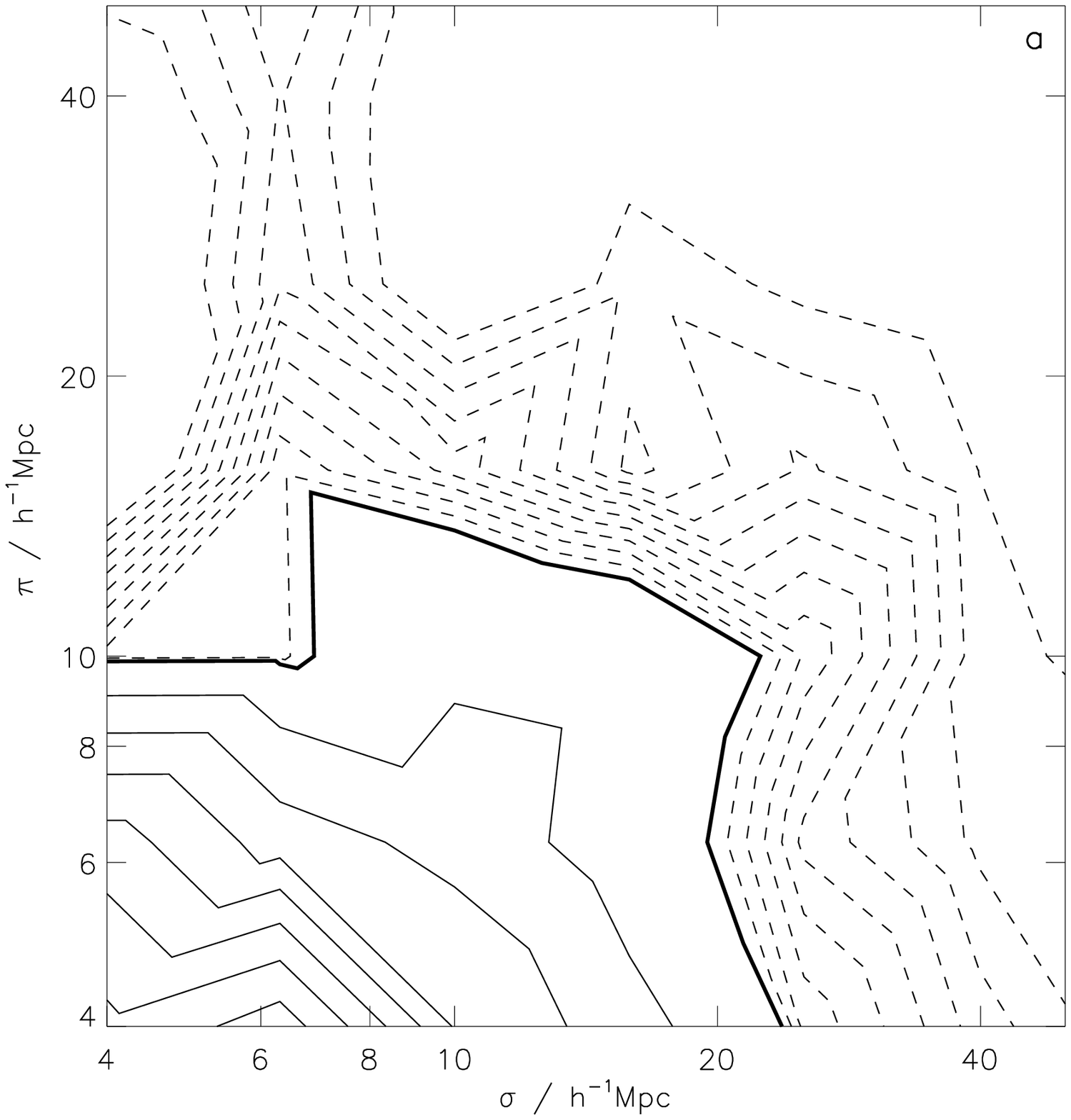}} &
{\epsfxsize=5.7truecm \epsfysize=5.truecm \epsfbox[10 10 550 560]{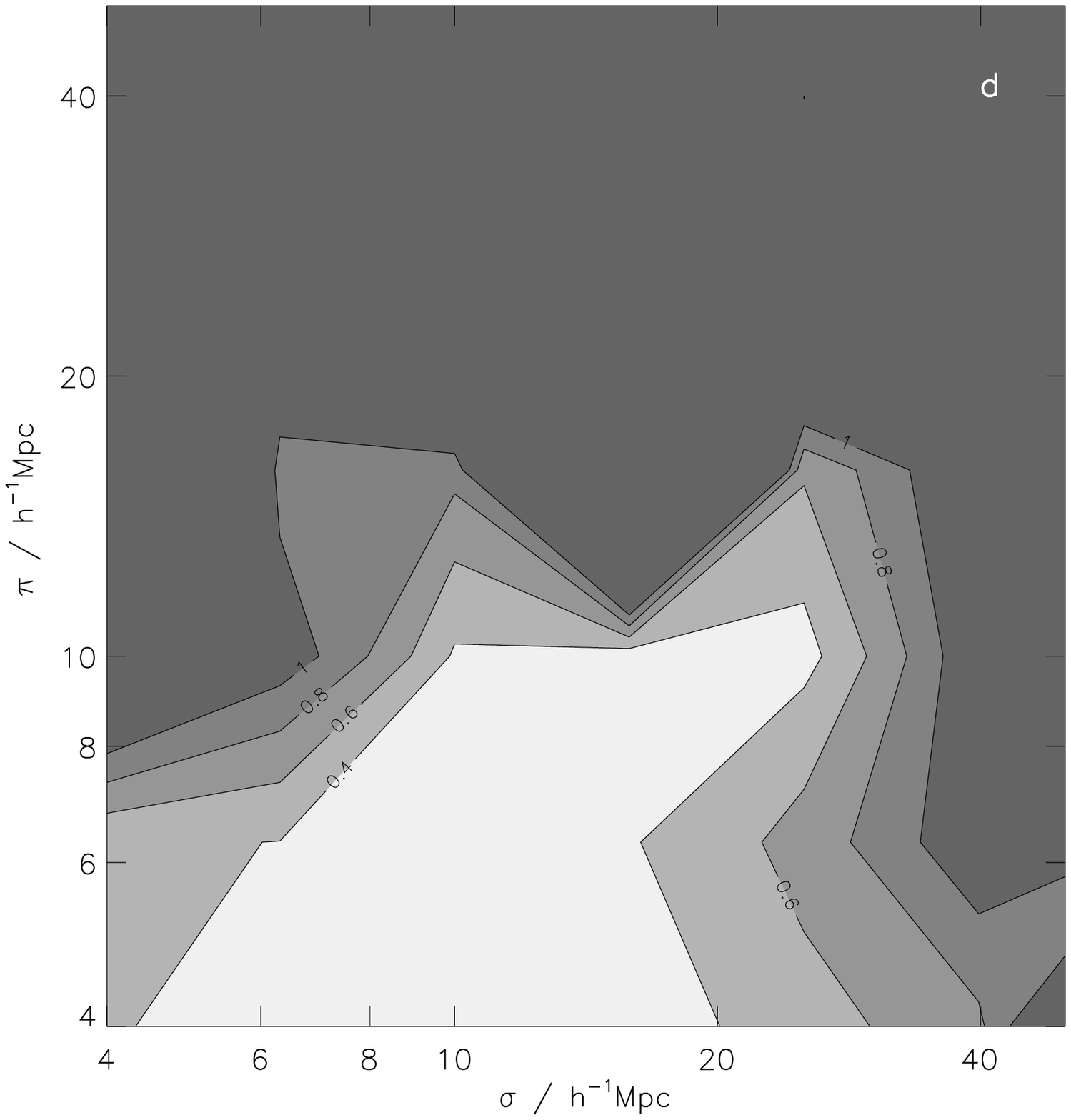}} &
{\epsfxsize=5.7truecm \epsfysize=5.truecm \epsfbox[10 10 550 560]{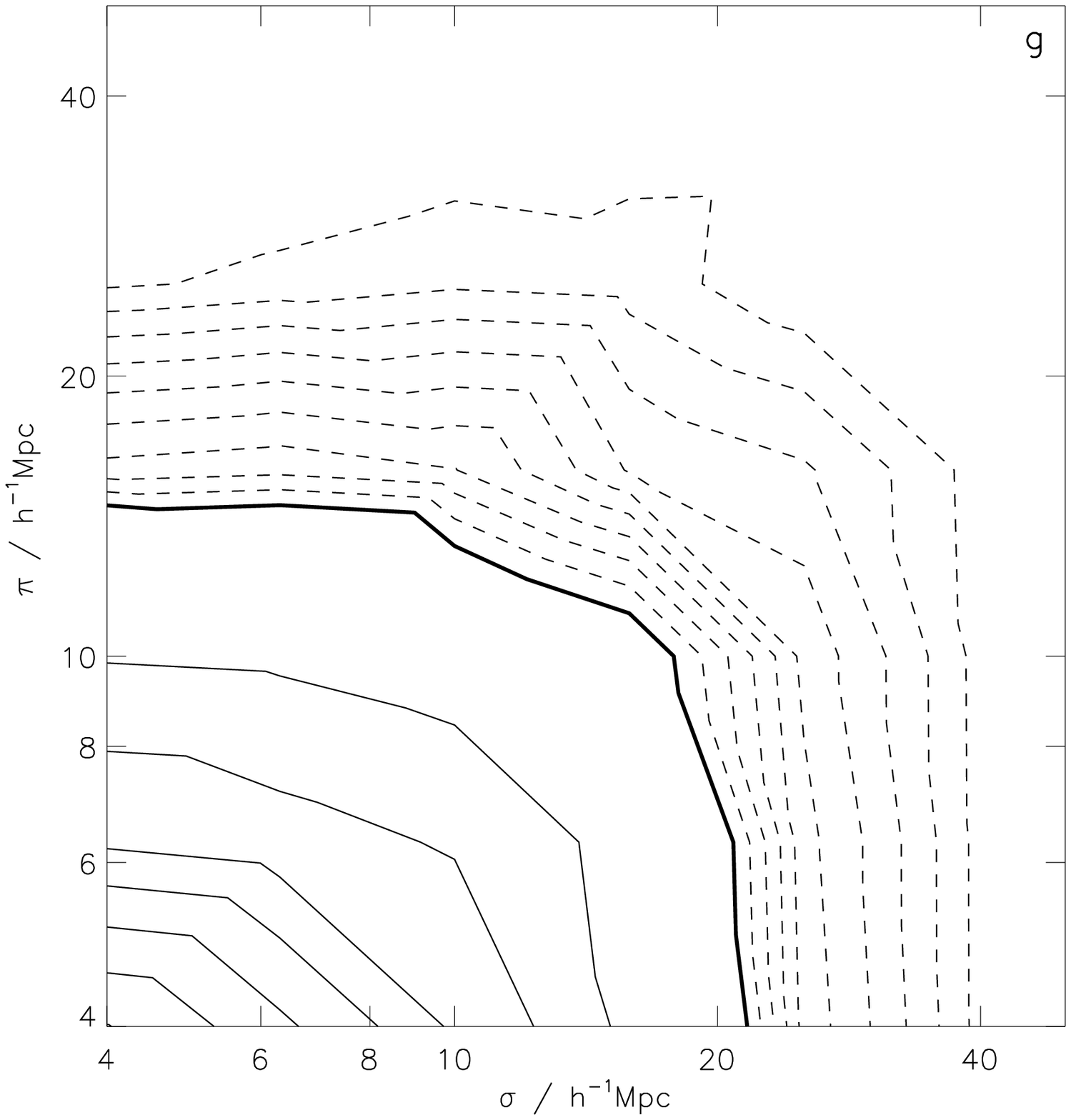}} \\
{\epsfxsize=5.7truecm \epsfysize=5.truecm \epsfbox[10 10 550 560]{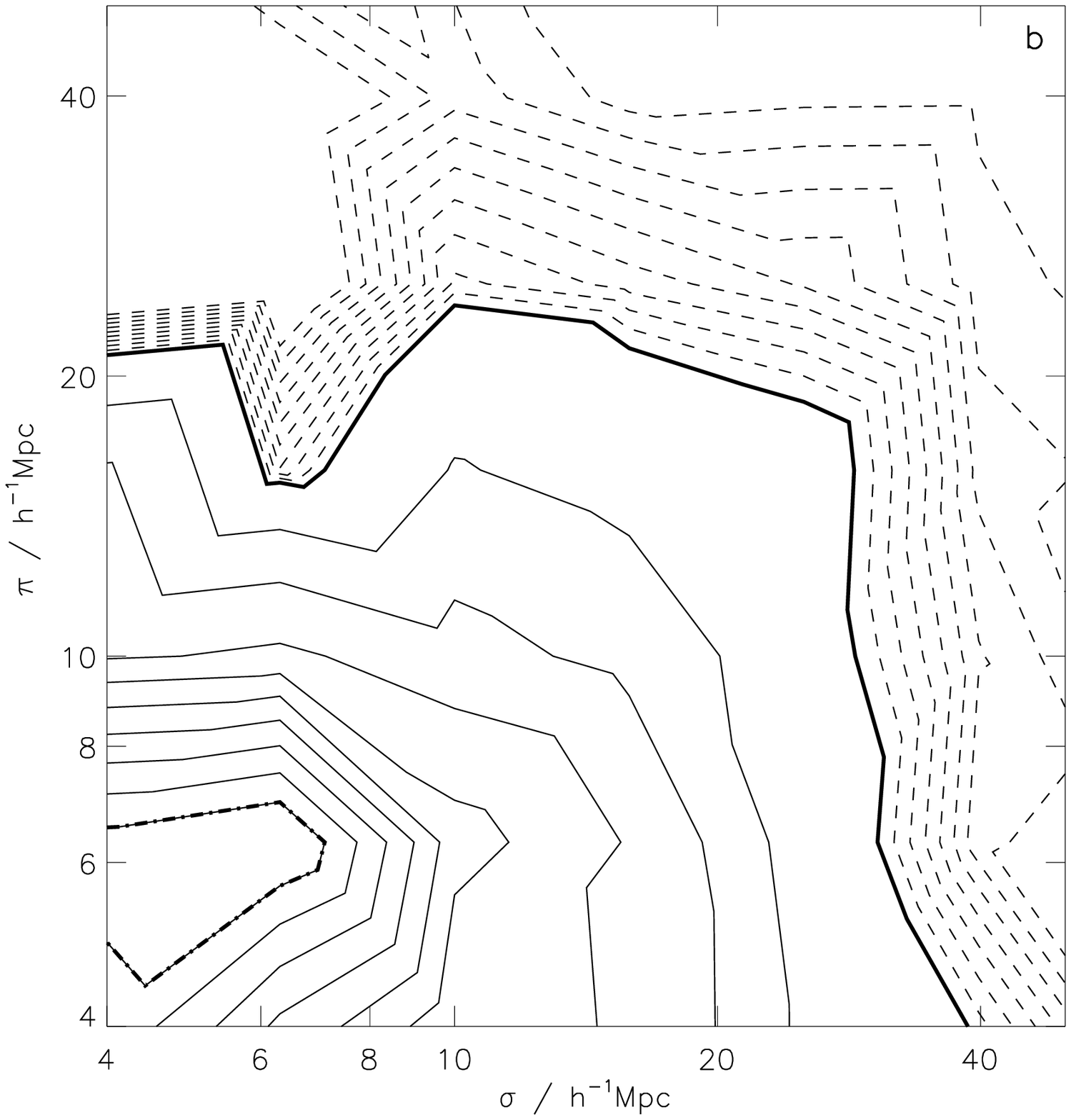}} &
{\epsfxsize=5.7truecm \epsfysize=5.truecm \epsfbox[10 10 550 560]{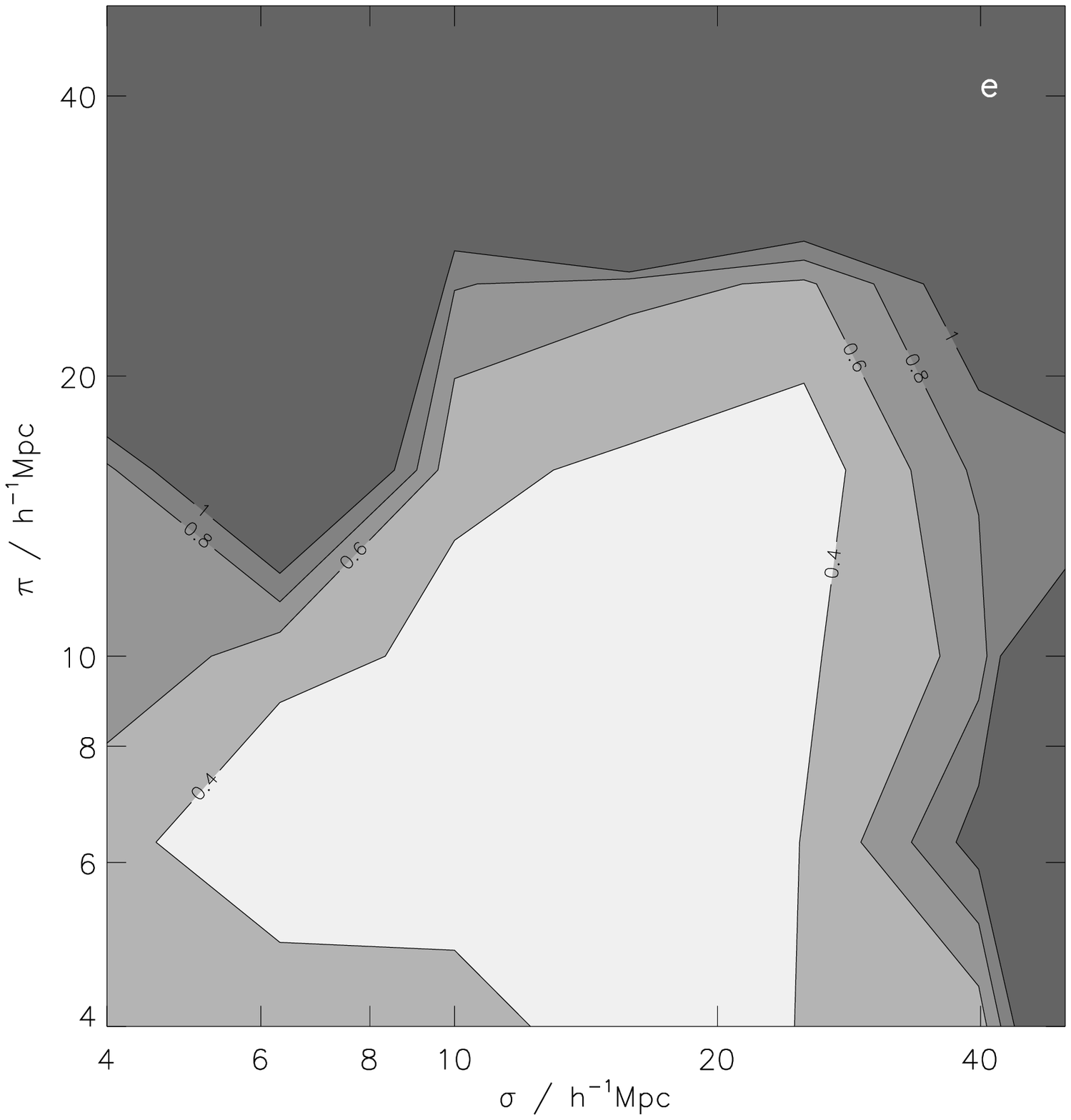}} &
{\epsfxsize=5.7truecm \epsfysize=5.truecm \epsfbox[10 10 550 560]{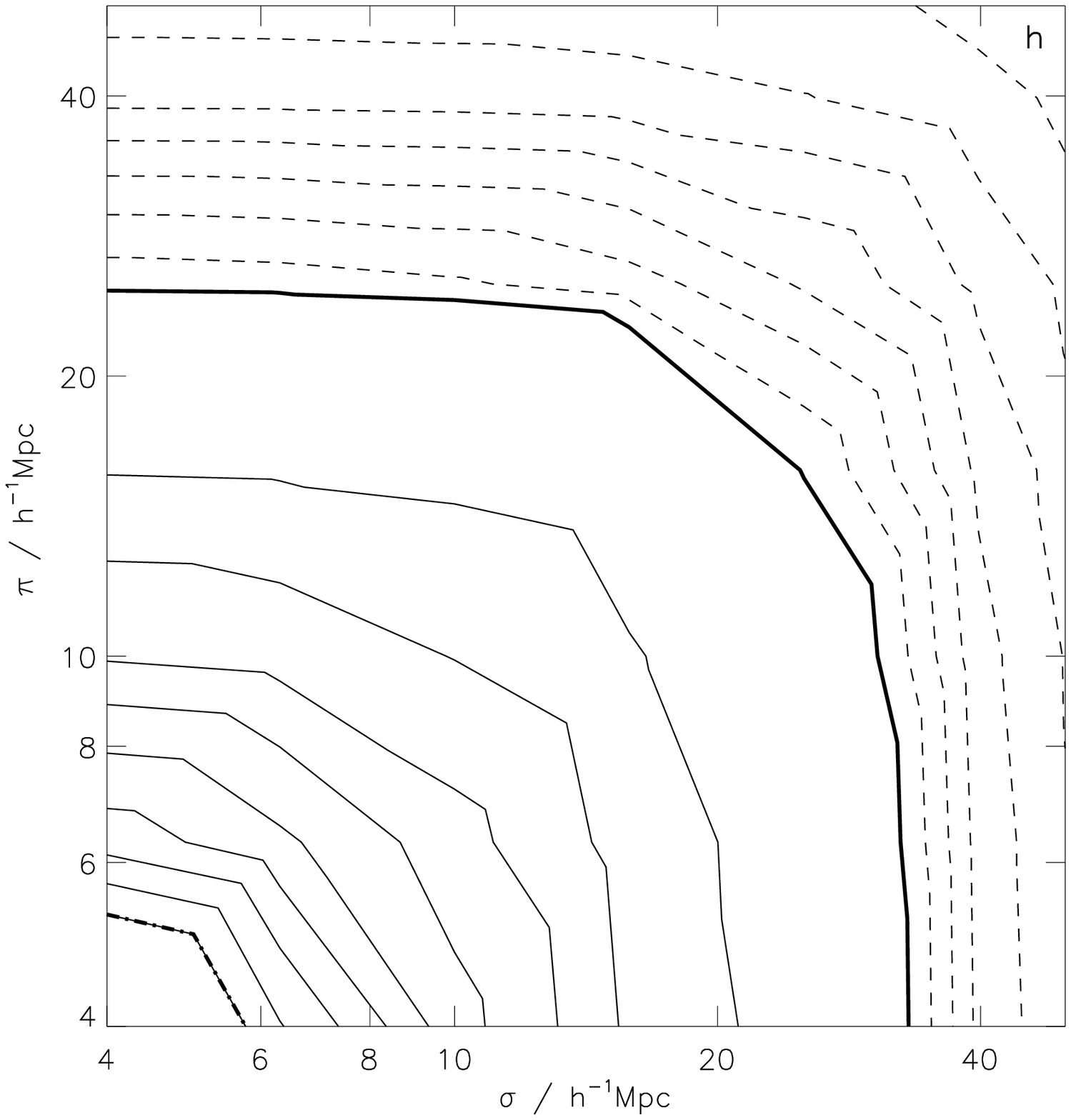}} \\
{\epsfxsize=5.7truecm \epsfysize=5.truecm \epsfbox[10 10 550 560]{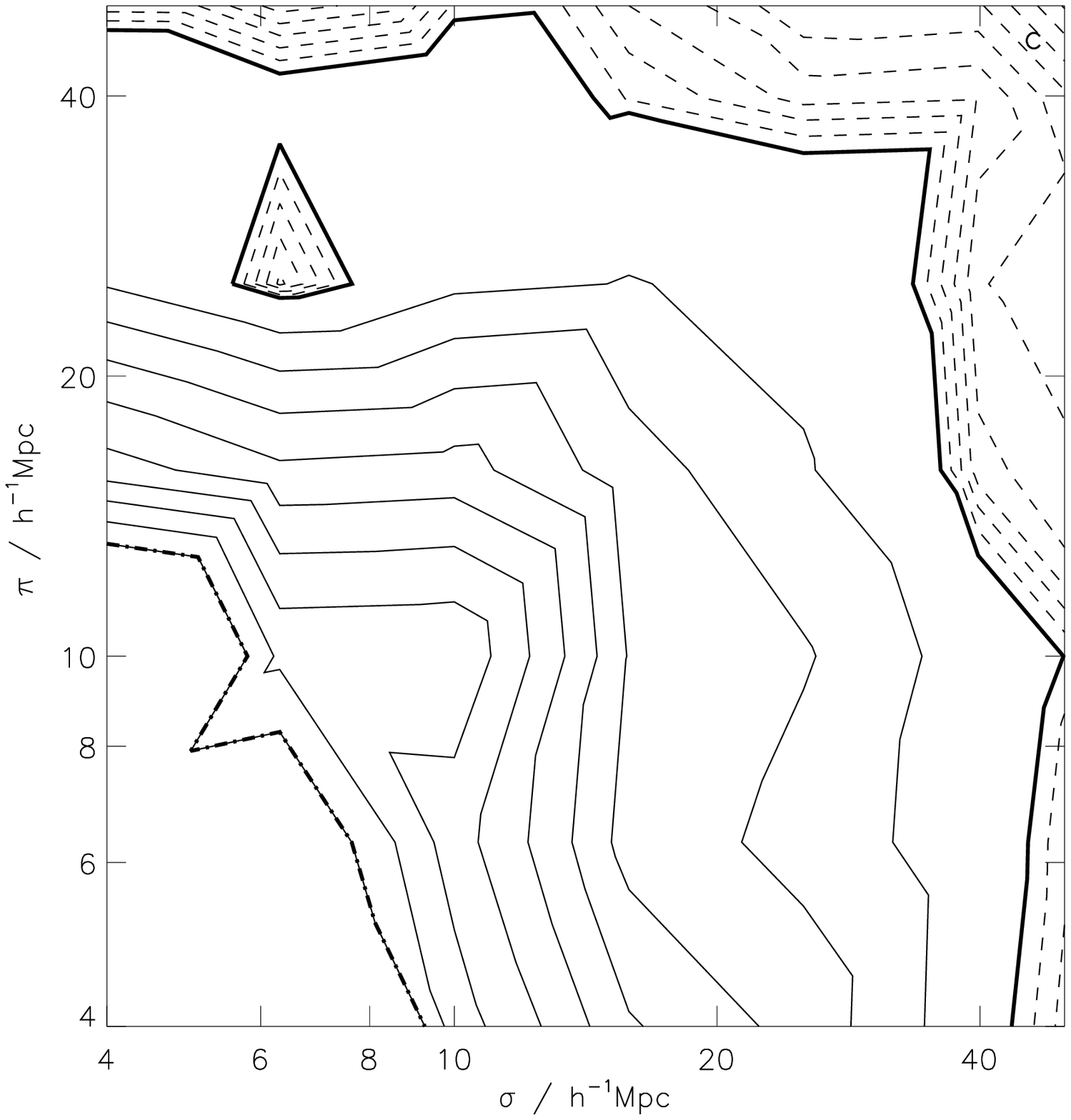}} &
{\epsfxsize=5.7truecm \epsfysize=5.truecm \epsfbox[10 10 550 560]{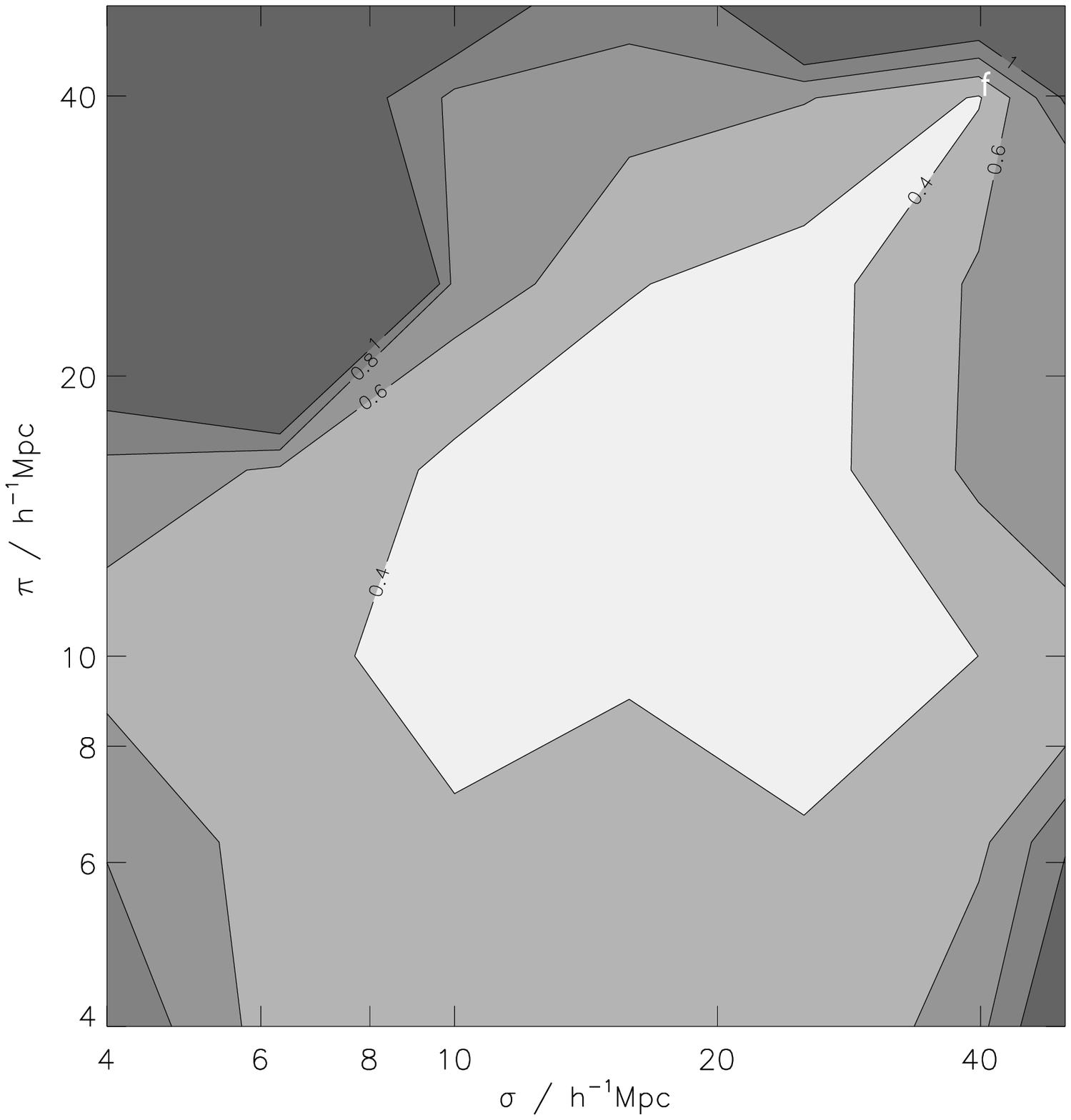}} &
{\epsfxsize=5.7truecm \epsfysize=5.truecm \epsfbox[10 10 550 560]{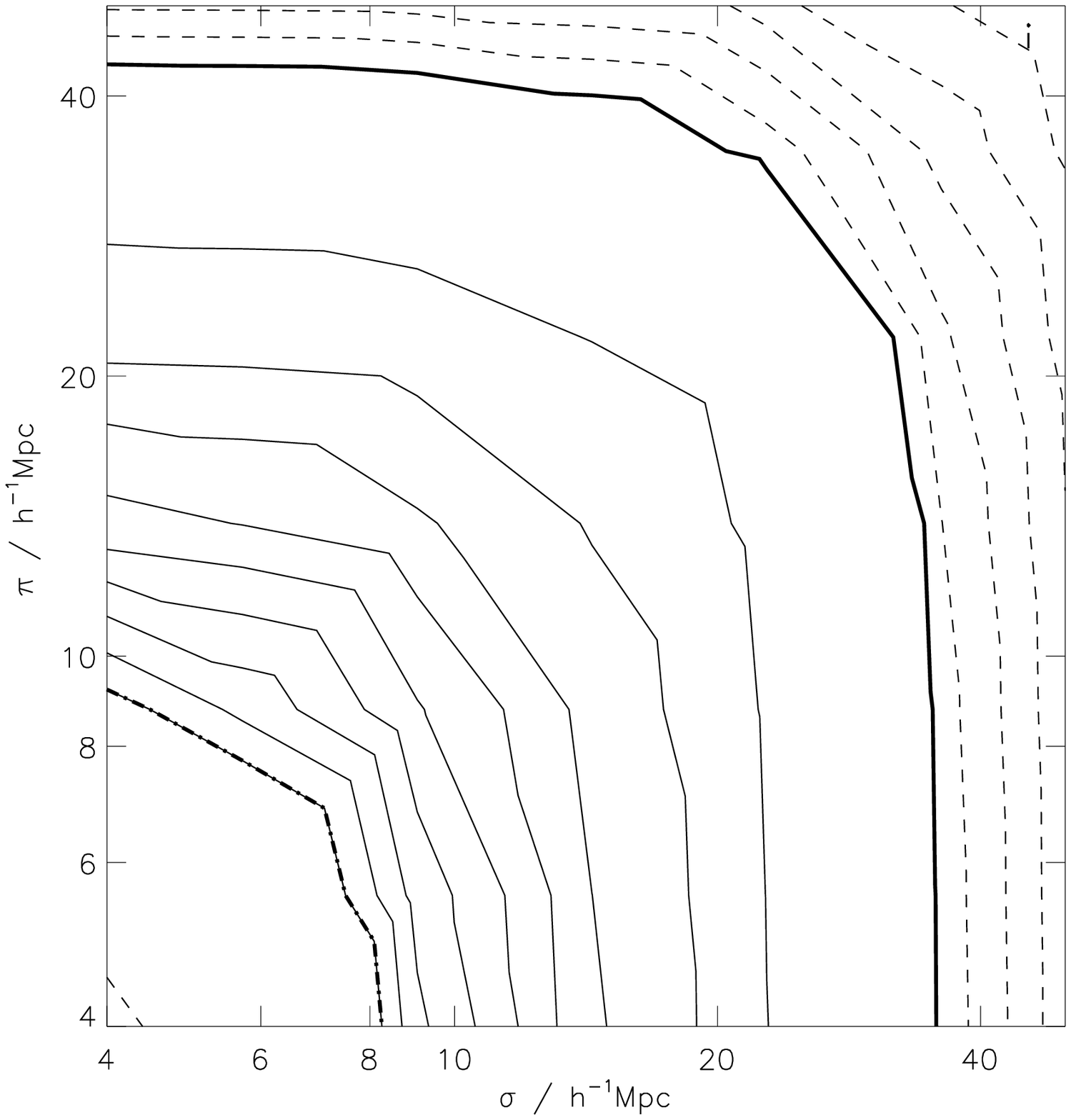}} \\
\end{tabular}
\caption
{The left hand plots show $\xi(\sigma,\pi)$ measured from the mock catalogues. The assumed cosmology is $\Omega_{\rm m}(0)$=1, $\Omega_{\Lambda}(0)$=0 (top row), $\Omega_{\rm m}(0)$=0.3, $\Omega_{\Lambda}(0)$=0.7 (middle row) and $\Omega_{\rm m}(0)=0$, $\Omega_{\Lambda}(0)=1$ (bottom row). Each strip contains 12,500 mock QSOs and the average over the two outer strips is shown. The solid, bold contour shows $\xi=0.1$, with the other solid lines increasing logarithmically from it with 0.1 interval and the bold, dot-dashed contour shows $\xi=1$. See the text for the value of the other lines. The centre plots show the fractional errors on $\xi(\sigma,\pi)$. The white area shows where $\xi(\sigma,\pi)$ is measured to better than 40\% and with each contour the fractional error increase by 20\%. The right hand plots show models of $\xi(\sigma,\pi)$. The values of the velocity dispersion, $\beta_{QSO}(\bar{z})$ and the test cosmology are taken from the {\it Hubble Volume} simulation. }
\label{fig:HVsigpi}
\end{centering}
\end{figure*}

Does the model reproduce $\xi(\sigma,\pi)$ measured from the mock QSO catalogues? The left hand plots in Figure \ref{fig:HVsigpi} show $\xi(\sigma,\pi)$ measured from the mock catalogues. We show the fractional errors on $\xi(\sigma,\pi)$ in the middle plots --- the lighter the shading, the smaller the errors. The areas of the $\xi(\sigma,\pi)$ diagram where the fractional errors are the smallest are the areas where most of the differentiation between different models of $\xi(\sigma,\pi)$ can be made. The right hand plots show the model $\xi(\sigma,\pi)$. Three different cosmologies have been assumed for the simulation and the models, $\Omega_{\rm m}(0)$=1, $\Omega_{\Lambda}(0)$=0 (shown in panels a, d and g), $\Omega_{\rm m}(0)$=0.3, $\Omega_{\Lambda}(0)$=0.7 (panels b, e and h), which is also the simulation cosmology, and $\Omega_{\rm m}(0)=0$, $\Omega_{\Lambda}(0)=1$ (panels c, f and i). The bold, solid lines show $\xi$=0.1, the dot dashed bold lines show $\xi$=1. The solid lines increase from $\xi$=0.1 in steps of 0.1, the dashed lines decrease from $\xi=0.1$ in steps of 0.01 and the dot-dashed lines increase from $\xi$=1 in steps of 1. The values of the velocity dispersion ($<w_z^2>^{1/2}$=400 km s$^{-1}$), $\beta_{QSO}(\bar{z})$ (=0.36) and test cosmology in the models are the values measured from the Hubble Volume simulation at the average redshift.

The measurement of $\xi(\sigma,\pi)$ from the Hubble Volume simulation is quite noisy. However, particularly if the bold, solid line ($\xi=0.1$) is considered, the effects of the different assumed cosmologies can be seen. This line in panel a) is far more squashed in the $\pi$ direction than in panel b) and elongation is seen in panel c) as compared to panel b). This is seen in the models too. The solid line in panel d) appears more squashed in the $\pi$ direction than the lines in panel e) and the lines in panel f) appear more elongated in the $\pi$ direction than the lines in panel e). For each choice of assumed cosmology, the models match the simulation to within $\sim 1 \sigma$.

\section{Determining the Underlying (Simulation) Cosmology}
\label{sec:true}

We wish to use geometric distortions in the measurement of $\xi(\sigma,\pi)$ to constrain the cosmological parameters $\Omega_{\rm m}$ and $\Omega_{\Lambda}$. We apply the method to QSOs as it is only at high redshifts that the geometric distortions become apparent. Current, wide-angled galaxy redshift surveys only probe the universe out to $z \sim 0.1-0.2$ where an incorrect choice of cosmology leads to a 10\% uncertainty in separations for flat cosmologies. This produces only a small distortion in the clustering pattern which is very hard to seperate from the effects of bulk flows. However, if clustering is measured over a wider range of redshifts, the effects of assuming the wrong cosmology become large, as can be seen in Figure \ref{fig:HVsigpi}.

We constrain cosmology by comparing $\xi(\sigma,\pi)$ measured from the data, or here the {\it Hubble Volume}, to our models. There are a number of free parameters in the models, the test cosmology, $\beta_{QSO}(\bar{z})$ and the small scale peculiar velocities, $<w_z^2>^{1/2}$. We fix the assumed cosmology to be either EdS or the $\Lambda$ cosmology to show that the results are independent of this choice.

\begin{figure} 
\begin{centering}
\begin{tabular}{c}
{\epsfxsize=5.5truecm \epsfysize=5truecm \epsfbox[10 10 550 560]{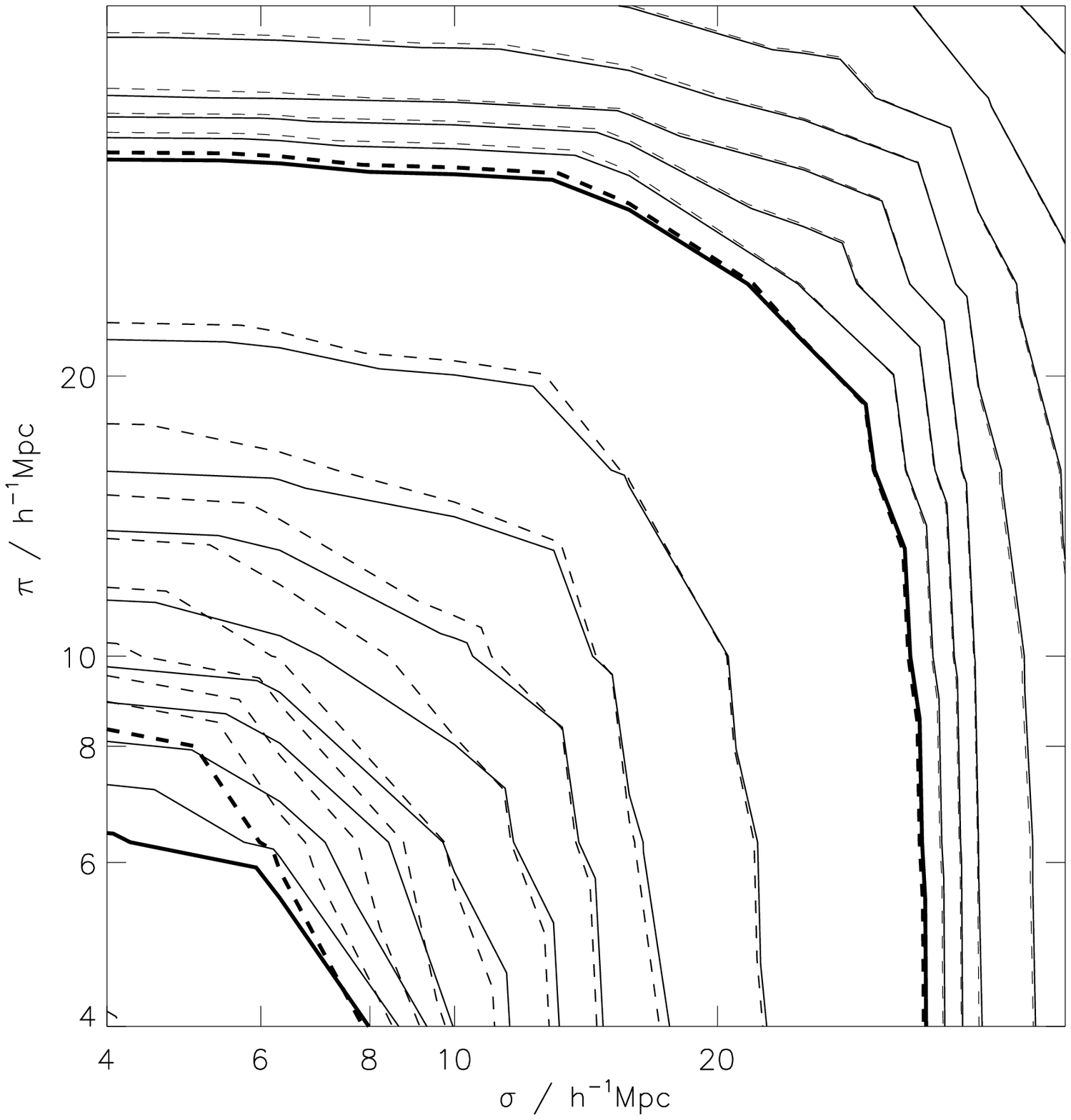}} \\
{\epsfxsize=5.5truecm \epsfysize=5truecm \epsfbox[10 10 550 560]{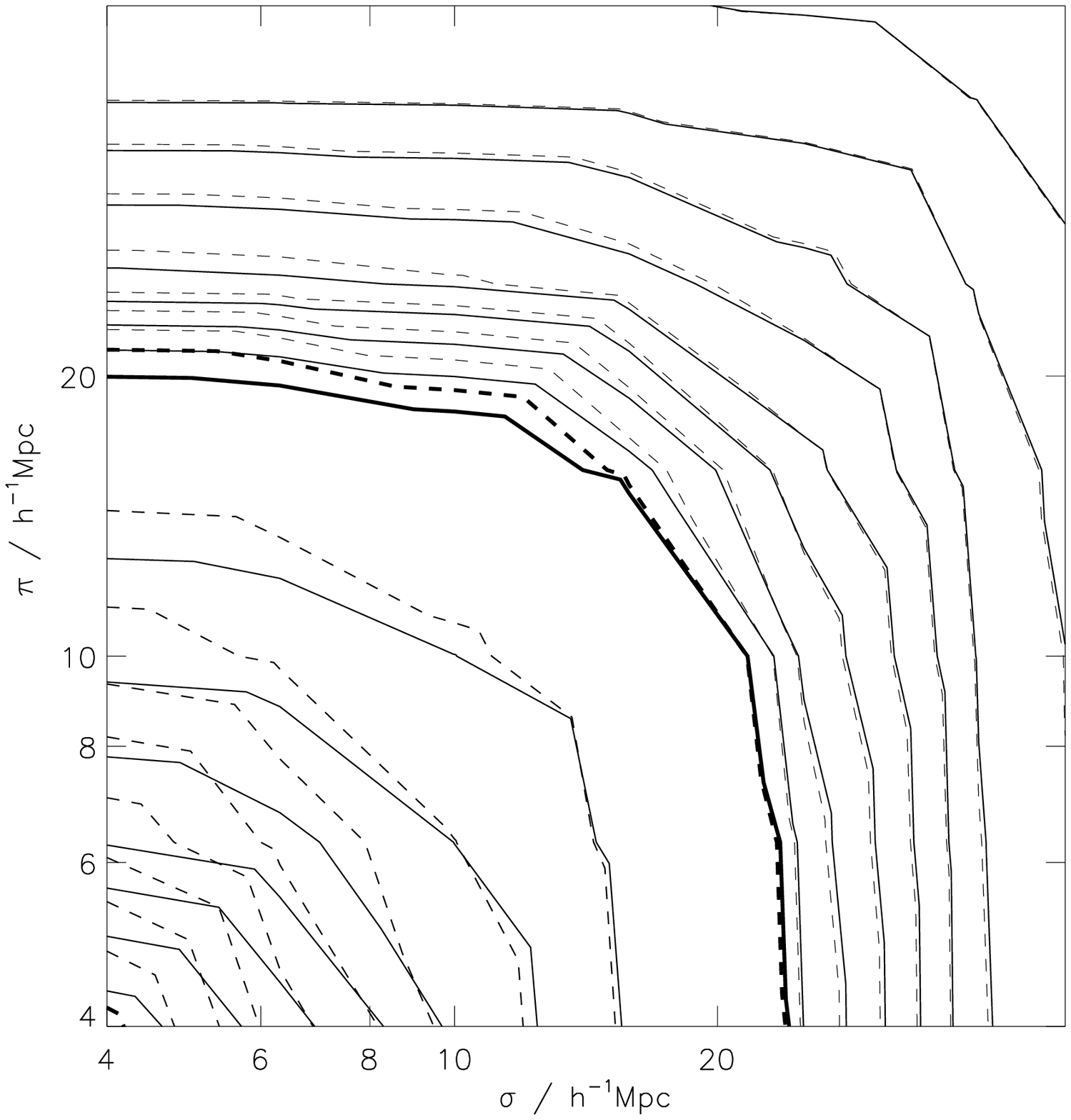}} \\
\end{tabular}
\caption {Two sets of models with different values of the small scale peculiar velocities. The top plot shows models with test cosmology $\Omega_{\rm m}(0)$=1., $\Omega_{\Lambda}(0)$=0., $\beta_{QSO}(\bar{z})$=0.36 and an assumed $\Lambda$ cosmology. The solid lines have $<w_z^2>^{1/2}$= 50 km s$^{-1}$ and the dashed lines have $<w_z^2>^{1/2}$= 1000 km s$^{-1}$. The lower plot shows models with test cosmology $\Omega_{\rm m}(0)$=0.3, $\Omega_{\Lambda}(0)$=0.7, $\beta_{QSO}(\bar{z})$=0.36 and an assumed $\Lambda$ cosmology. Again the solid lines have $<w_z^2>^{1/2}$= 50 km s$^{-1}$ and the dashed lines have $<w_z^2>^{1/2}$= 1000 km s$^{-1}$. Apart from on the smallest scales, there is little difference between the two models considering the large difference between the two velocities and the error contours shown in Figure \ref{fig:HVsigpi}.}
\label{fig:velsdemo}
\end{centering}
\end{figure}

As we compare the models and the simulation on scales greater than 4$h^{-1}$Mpc in both the $\pi$ and $\sigma$ direction, we find that the modeling of $\xi(\sigma,\pi)$ is fairly insensitive to the value of the small scale velocity dispersion. This is shown in Figure \ref{fig:velsdemo}. Here we show two different sets of models, the top plot shows models with test cosmology $\Omega_{\rm m}(0)$=1, $\Omega_{\Lambda}(0)$=0, $\beta_{QSO}(\bar{z})$=0.36 and the $\Lambda$ cosmology assumed. The solid lines have $<w_z^2>^{1/2}=50$ km s$^{-1}$ and the dashed lines have $<w_z^2>^{1/2}=1000$ km s$^{-1}$. The lower plot shows models with test cosmology $\Omega_{\rm m}(0)$=0.3, $\Omega_{\Lambda}(0)$=0.7, $\beta_{QSO}(\bar{z})$=0.36 and the $\Lambda$ cosmology assumed. Again the solid lines have $<w_z^2>^{1/2}=50$ km s$^{-1}$ and the dashed lines have $<w_z^2>^{1/2}=1000$ km s$^{-1}$. Considering the wide range of velocities covered by these two models, there is little difference between them, apart from on the smallest $\sigma$ scales where the errors on $\xi(\sigma,\pi)$ from the mock catalogues are large anyway (see Figure \ref{fig:HVsigpi}). The models with both choices of $<w_z^2>^{1/2}$ still match the simulation $\xi(\sigma,\pi)$ to within 1$\sigma$. This is because most of the differentiation between different models of $\xi(\sigma,\pi)$ occurs on larger scales. We therefore fix $<w_z^2>^{1/2}$= 400 km s$^{-1}$ in the models when we are finding the {\it Hubble Volume} cosmology.

\begin{figure} 
\begin{centering}
\begin{tabular}{c}
{\epsfxsize=7truecm \epsfysize=7truecm \epsfbox[30 150 550 670]{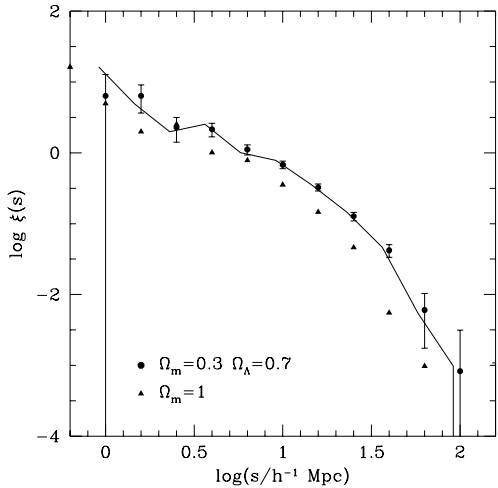}} \\
\end{tabular}
\caption{The redshift space correlation function of the mock catalogues estimated assuming the $\Lambda$ cosmology (circles) and EdS (triangles). The line shows the correlation function measured assuming EdS corrected to the $\Lambda$ cosmology with the relation between $s$ and $s^{\prime}$ given in equation \ref{eq:rscale}. This line agrees well with the circles over the range of scale 3$< r <$50$h^{-1}$Mpc, showing that a correlation function measured in one cosmology can be translated into another. }
\label{fig:xicosmo}
\end{centering}
\end{figure}

As the effects of small scale peculiar velocities have little impact on models of $\xi(\sigma,\pi)$ on scales larger than 4$h^{-1}$Mpc, they should also have little effect on the correlation function $\xi(s)$. We have tested this on the {\it Hubble Volume} simulation by comparing the real and redshift space correlation functions of the mass. We find that on scales greater than $\sim 3h^{-1}$Mpc the small scale peculiar velocities have negligible effect on the shape of the redshift space mass correlation function. It is therefore possible to find the real space correlation function self-consistently from a redshift space correlation function for each test value of $\beta$ via the formula of \citeasnoun{Kaiser87}

\begin{equation}
\xi(r) = \frac{\xi(s)} {\left\{ 1 + \frac{2}{5}\beta + \frac{1}{5}\beta^2 \right\}},
\label{eq:kaiser}
\end{equation}
where $r$ and $s$ denote the correlation function in real and redshift space respectively and $\beta = \Omega_{\rm m}(0)^{0.6}/b$. This is measured in the assumed cosmology but, to create the model $\xi(\sigma,\pi)$, we need the real space correlation function measured in the test cosmology. First we translate the redshift space correlation function, $\xi(s)$, in the test cosmology to the assumed cosmology, $\xi(s^{\prime})$ by scaling $s$ to $s^{\prime}$ using
\begin{equation}
s^2 = \sigma^2 + r_z^2 
\end{equation}
\noindent and 
\begin{equation}
s^{\prime 2} = (\sigma/f_{\perp})^2 + (r_z/f_{\parallel})^2
\label{eq:rscale}
\end{equation}
then we apply equation \ref{eq:kaiser}.
The terms $f_{\perp}$ and $f_{\parallel}$ are defined in equation \ref{eq:sigscale} and \ref{eq:piscale} as well as in \citeasnoun{BPH96}. The accuracy of converting the two-point correlation function from one cosmology into another is demonstrated in Figure \ref{fig:xicosmo}. The symbols show the redshift space correlation function measured from the {\it Hubble Volume} with EdS assumed (triangles) and the $\Lambda$ cosmology assumed (circles). The line shows the correlation function measured in the EdS cosmology but translated into the $\Lambda$ cosmology. There is excellent agreement between the two over a wide range of scales. This is a further consistency check that measuring $f_{\perp}$ and $f_{\parallel}$ at the average redshift of the Survey is an adequate approximation to make. An alternative but more time consuming method would be to measure the redshift space correlation function in each of the test cosmologies directly and then apply equation \ref{eq:kaiser} to obtain the real space correlation function in the test cosmology.

We now just have two free parameters, the test cosmology and the value of $\beta_{QSO}(\bar{z})$. We run a grid of models and find which choice of test cosmology and $\beta_{QSO}(\bar{z})$ give the best fit to the data. 

\subsection{The Fitting Procedure}

The fitting procedure that we adopt to find which test cosmology matches the simulation cosmology is as follows:

\noindent 1) Pick an assumed cosmology (here either EdS or the $\Lambda$ cosmology)

\noindent 2) Calculate $\xi(s)$ and $\xi(\sigma,\pi)$ from the data (the 2QZ Survey or the mock catalogues) using the assumed cosmology.

\noindent 3) Pick a value for the test $\beta_{QSO}(\bar{z})$ at the average redshift of the survey.

\noindent 4) Pick a value of the present day test $\Omega_{\rm m}(0)$ and calculate $\Omega_{\rm m}(\bar{z})$ at $\bar{z}$=1.4 via $\Omega_{\rm m}(\bar{z}) = \Omega_{\rm m}(0)(1+\bar{z})^3/[\Omega_{\rm m}(0) (1+\bar{z})^3 + \Omega_{\Lambda}(0)(0)]$.

\noindent 5) Calculate the bias $b_{QSO}(\bar{z}) = \Omega_{\rm m}(0)^{0.6}(\bar{z})/ \beta_{QSO}(\bar{z}) $.

\noindent 6) Translate the redshift space correlation function, $\xi(s)$, measured in the assumed cosmology to a real space correlation function, $\xi(r^{\prime})$, in the test cosmology using equation \ref{eq:rscale} and \ref{eq:kaiser}.

\noindent 7) Generate the model $\xi(\sigma,\pi)$.

\noindent 8) Translate the model $\xi(\sigma,\pi)$ from the test cosmology into the assumed cosmology using equations \ref{eq:piscale} and \ref{eq:sigscale}.

\noindent 9) Calculate how well the model $\xi(\sigma,\pi)$ fits the data $\xi(\sigma,\pi)$ via the $\chi^2$ statistic, using the Poisson errors from the data $\xi(\sigma,\pi)$ measured in the assumed cosmology.

\noindent 10) Go back to 3) using a different test cosmology and $\beta_{QSO}(\bar{z})$.

When the parameters $\beta_{QSO}(\bar{z})$ and $\Omega_{\rm m}(0)$ match those of the underlying cosmology (or in this case the simulation cosmology), the value of $\chi^2$ should be minimised. We fit the model $\xi(\sigma,\pi)$ to the {\it Hubble Volume} $\xi(\sigma,\pi)$ over the range of scales 4$<\sigma,\pi<$40$h^{-1}$ Mpc. This is to ensure that any non-linear effects are small and that the errors on $\xi(\sigma,\pi)$ do not dominate the actual value of $\xi(\sigma,\pi)$.

The error contours, shown in Figure \ref{fig:HVres} are found as follows. The number of degrees of freedom, $\nu$, is the number of bins in which the model $\xi(\sigma,\pi)$ is fitted to the data $\xi(\sigma,\pi)$ minus the number of free parameters (this is 2, $\Omega_{\rm m}(0)$ and $\beta_{QSO}(\bar{z})$). We have checked the error contours using Monte Carlo simulations and find the two methods to give comparable results.

\section{Predicted Results for the 2QZ Survey}
\label{sec:reslts}

\begin{figure} 
\begin{centering}
\begin{tabular}{c}
{\epsfxsize=8truecm \epsfysize=8truecm \epsfbox[10 10 550 560]{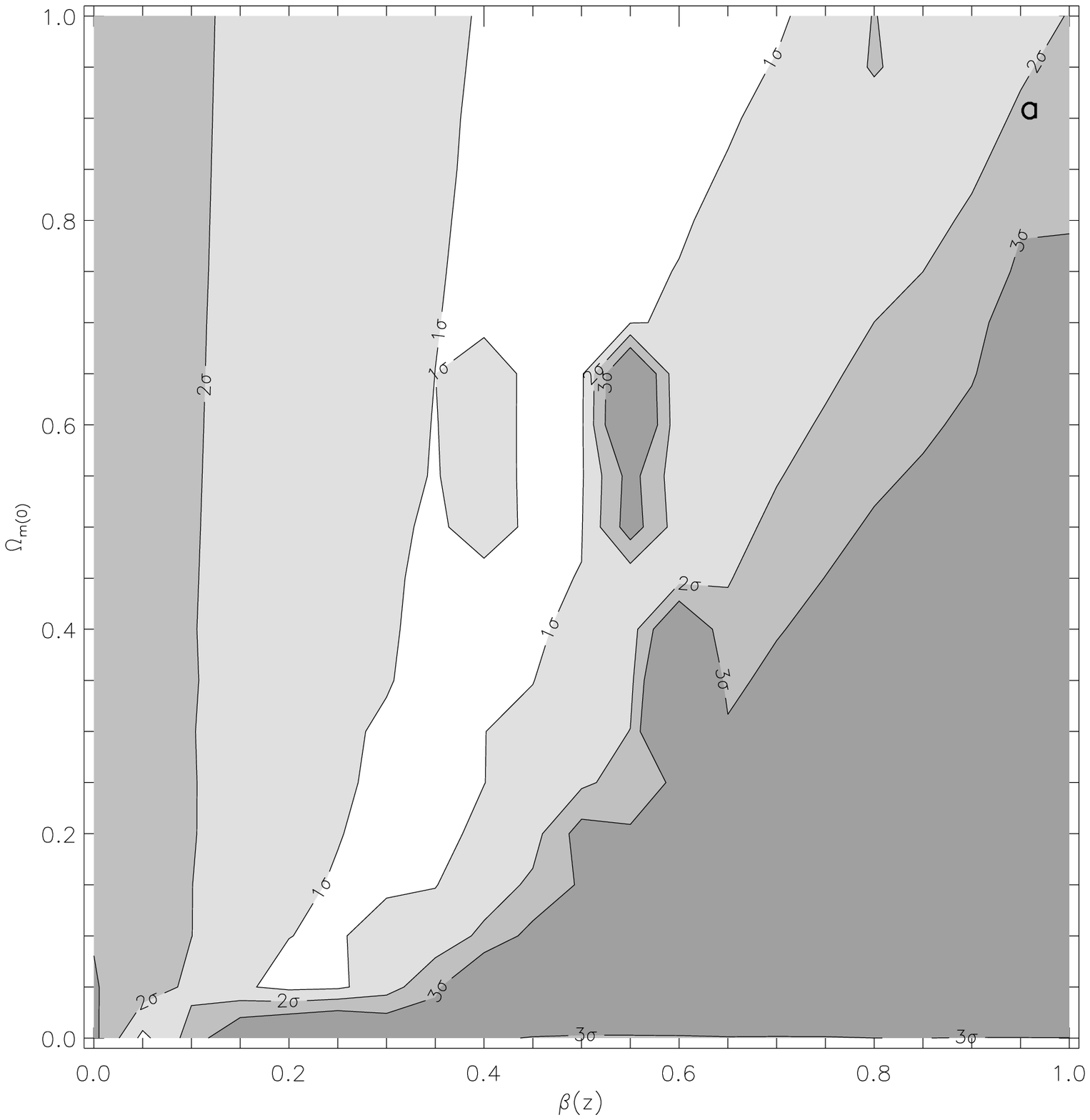}} \\
{\epsfxsize=8truecm \epsfysize=8truecm \epsfbox[10 10 550 560]{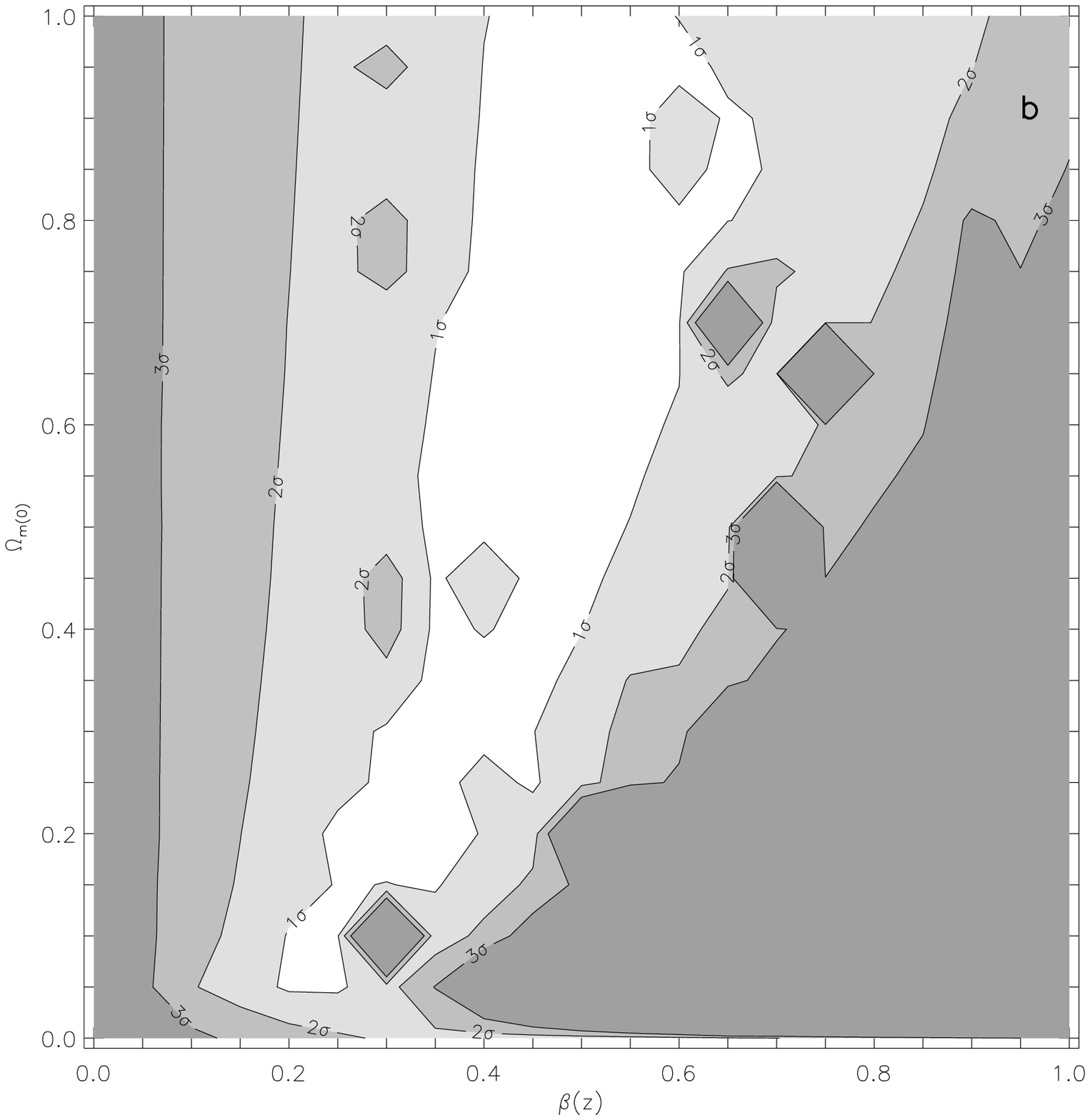}} \\
\end{tabular}
\caption
{Fitting $\xi(\sigma,\pi)$ from the {\it Hubble Volume}. The top panel (a) has EdS as the assumed cosmology and the lower panel (b) has the $\Lambda$ cosmology assumed. In both cases the simulation cosmology, $\Omega_{\rm m}(0)$=0.3, $\Omega_{\Lambda}(0)$=0.7 and $\beta_{QSO}(\bar{z})$=0.36 is contained within the 1$\sigma$ contour, although little constraint on $\Omega_{\rm m}(0)$ is possible from the mock 2QZ Survey catalogues.}
\label{fig:HVres}
\end{centering}
\end{figure}

The results of fitting the $\xi(\sigma,\pi)$ models to $\xi(\sigma,\pi)$ measured from the mock catalogues are given in Figure \ref{fig:HVres}. The upper panel (a) shows the results when the assumed cosmology is EdS and the lower panel (b) is for the $\Lambda$ cosmology. We show the results for two different assumed cosmologies to show they are not dependent on this choice. In both cases, the simulation cosmology, $\Omega_{\rm m}(0)$=0.3, $\Omega_{\Lambda}(0)$=0.7 and $\beta_{QSO}(\bar{z})=0.36$ is contained within the 1$\sigma$ contour. However, the results do not place a strong constraint on $\Omega_{\rm m}(0)$ as there is a degeneracy between $\Omega_{\rm m}(0)$ and $\beta_{QSO}(\bar{z})$. However, the value of $\beta_{QSO}(\bar{z})$ with $\bar{z}=1.4$ should be fairly well constrained from the 2QZ Survey. This does at least place a joint constraint on cosmology and the QSO-mass bias.

\begin{figure} 
\begin{centering}
\begin{tabular}{c}
{\epsfxsize=8truecm \epsfysize=8truecm \epsfbox[10 10 550 560]{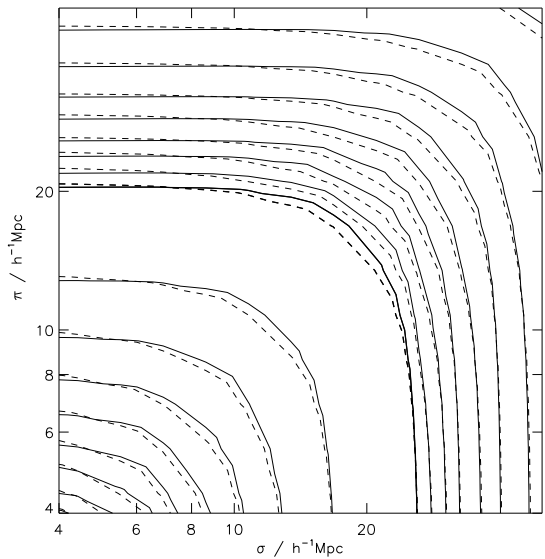}} \\
\end{tabular}
\caption
{The dashed lines show a model with test cosmology $\Omega_{\rm m}(0)$=0.3 and $\Omega_{\Lambda}(0)$=0.7 with a value of $\beta_{QSO}(\bar{z})$=0.35. The solid lines show a model with test cosmology $\Omega_{\rm m}(0)$=1 and $\beta_{QSO}(\bar{z})$=0.5. The difference between the two models is small. The $\Lambda$ cosmology is assumed in both cases.}
\label{fig:demo}
\end{centering}
\end{figure}

The reason for this degeneracy is that the errors on $\xi(\sigma,\pi)$ from the mock catalogues are fairly large. Shown in Figure \ref{fig:demo} are two different $\xi(\sigma,\pi)$ models. One has $\Omega_{\rm m}(0)$=0.3, $\Omega_{\Lambda}(0)$=0.7 and a value of $\beta_{QSO}(\bar{z})$=0.35 (dashed lines) and another model has $\Omega_{\rm m}(0)$=1, $\Omega_{\Lambda}(0)$=0 and $\beta_{QSO}(\bar{z})$=0.5 (solid lines). All the other parameters (the assumed $\Lambda$ cosmology and the small scale peculiar velocities of $<w_z^2>^{1/2}$=400 km s$^{-1}$ etc) are the same in each model. These models both provide a good fit to the mock catalogues from the {\it Hubble Volume} (see Figure \ref{fig:HVres}) which is not surprising as the two models have similar shape and amplitude. Very small errors on $\xi(\sigma,\pi)$ are needed to detect the small differences between these models. 

The degeneracy between $\Omega_{\rm m}(0)$ and $\beta_{QSO}(\bar{z})$ seen in Figure \ref{fig:HVres} does not lie in the direction that is perhaps intuitively expected. Consider the two assumed cosmologies adopted here, the two test cases shown in Figure \ref{fig:demo} and the simulation cosmology $\Omega_{\rm m}(0)$=0.3, $\Omega_{\Lambda}(0)$=0.7. If EdS is assumed then $\xi(\sigma,\pi)$ from the simulation is squashed (seen in Figure \ref{fig:HVsigpi}(a)), as is the model $\xi(\sigma,\pi)$ with test cosmology $\Omega_{\rm m}(0)$=0.3, $\Omega_{\Lambda}(0)$=0.7, due to the simulation and test cosmologies being different to the assumed cosmology. $\xi(\sigma,\pi)$ using $\Omega_{\rm m}(0)$=1, $\Omega_{\Lambda}(0)$=0 as the test cosmology is not compressed as the test cosmology matches the assumed cosmology. A higher value of $\beta_{QSO}(\bar{z})$ is therefore required to squash the model $\xi(\sigma,\pi)$ to match the simulation. 

If the $\Lambda$ cosmology is assumed, then no squashing due to cosmology occurs on the simulation $\xi(\sigma,\pi)$ (again seen in Figure \ref{fig:HVsigpi}(b)) or on the model $\xi(\sigma,\pi)$ with $\Omega_{\rm m}(0)$=0.3, $\Omega_{\Lambda}(0)$=0.7 as the test cosmology as in this case the simulation and test cosmologies match the assumed cosmology. However, if $\Omega_{\rm m}(0)$=1, $\Omega_{\Lambda}(0)$=0 is the test cosmology, then the model $\xi(\sigma,\pi)$ is elongated due to the different test and assumed cosmologies. Therefore, a larger value of $\beta_{QSO}(\bar{z})$ is again required to compensate for the elongation, allowing the model to match the simulation $\xi(\sigma,\pi)$. This is partly shown in Figure \ref{fig:velsdemo} as here we have models with the $\Lambda$ cosmology assumed and $\beta_{QSO}(\bar{z})$=0.35 (although the small scale peculiar velocities are 50 and 1000 km s$^{-1}$ rather than 400 km s$^{-1}$). The top plot has test cosmology $\Omega_{\rm m}(0)$=1, $\Omega_{\Lambda}(0)$=0 and the lower plot has test cosmology $\Omega_{\rm m}(0)$=0.3, $\Omega_{\Lambda}(0)$=0.7. Elongation is clearly seen in the top plot as compared to the bottom plot, showing that a higher value of $\beta_{QSO}(\bar{z})$ is required for the model with test cosmology $\Omega_{\rm m}(0)$=1, $\Omega_{\Lambda}(0)$=0 if the model is to match the simulation.

The results for determining the cosmology from the mock catalogues suggest that the cosmological parameters, $\Omega_{\rm m}(0)$ and $\Omega_{\Lambda}(0)$, will not be well constrained from the 2QZ Survey. However, the fits are only made to the mock 2QZ catalogues and there may be ways in which the constraints
from the final 2QZ may turn out better than these results suggest.

The correlation function from the 2QZ 10K Catalogue \cite{Scott00} has a powerlaw form over a slightly wider range of scales than the correlation function from the {\it Hubble Volume} simulations, which turns over to lower amplitude at small and large scales. Therefore, on small and large scales, the correlation function from 2QZ has a slightly higher amplitude than that from the {\it Hubble Volume}. This is mainly because the biasing scheme was chosen to match the clustering of the 2QZ correlation function on scales around 10$h^{-1}$Mpc but we also preserved the shape of the dark matter correlation function. The higher amplitude in the 2QZ correlation function on small and large scales helps in two ways. The extra large scale power may allow us to fit the models out to larger scales, increasing the number of bins over which we can compare the data and models. Also, the number of pairs found in the 2QZ Survey on small scales will be larger than the number found in the mock catalogues. This reduces the errors on small scales, hopefully giving tighter constraints on the models.

\section{Further Constraints on Cosmology and $\beta_{QSO}$({\tiny z})}
\label{sec:evo}

As shown in section \ref{sec:reslts}, the cosmological parameters, $\Omega_{\rm m}(0)$ and $\beta_{QSO}(\bar{z})$, will probably not be uniquely measured from the 2QZ Survey using this technique. However, there are other ways in which the cosmological parameters can be estimated from the 2QZ Survey and results can be combined to obtain stronger constraints on cosmology. The basic idea developed in this section is that the $\Omega_{\rm m}:\beta_{QSO}(\bar{z})$ degenerate set obtained from QSO clustering evolution is different from the $\Omega_{\rm m}:\beta_{QSO}(\bar{z})$ degenerate set obtained from analysing QSO space distortions and that by using these two constraints in combination, the degeneracy in these two parameters may be lifted.

The amplitude of QSO clustering can be measured at $z=1.4$ assuming any cosmology. In order to calculate the value of $\beta_{QSO}(1.4)$, we need to know the clustering amplitude of the mass at this redshift. To obtain this, we need to know the amplitude of the clustering of the mass at $z=0$, then linear perturbation theory allows us to trace the evolution of the mass clustering. 

However, the clustering of the mass at $z=0$ is not known. Here we use information from galaxy redshift surveys in order to determine the mass correlation function at $z=0$. The mass correlation function can be determined if the galaxy correlation function and $\beta_{\rm g}$ are known, assuming that bias is independent of scale.

Many measurements of $\beta_{\rm g}$ at $z=0$ have appeared in the literature, for example \citeasnoun{TE96}, \citeasnoun{Rat98c}, \citeasnoun{Hoyle99} and \citeasnoun{Outram}. The latest measurement of $\beta_{\rm g}$ comes from the 2dF Galaxy Redshift Survey and is 0.43$\pm$0.07 \cite{Peacock}. For each cosmology (again we only consider flat cosmologies) the value of the galaxy-mass bias can be found from $\beta_{\rm g}$, which in turn gives the value of the mass correlation function if the galaxy correlation function at $z=0$ is known, as shown below.

Rather than determining the value of the two-point correlation function at one particular point, we use the less noisy volume averaged two-point correlation function, $\bar{\xi}^s_g$. We assume that the galaxy correlation function can be approximated by a power law of the form $\xi^s_g=(s/6 h^{-1}$Mpc$)^{-1.7}$. $\bar{\xi}_g$ is then found via
\begin{equation}
\bar{\xi}^s_g = \frac{ \int_0 ^{20} 4 \pi s^2 (s/6)^{-1.7} ds}{\int_0 ^{20} 4 \pi s^2 ds}.
\end{equation}
By integrating out to 20$h^{-1}$Mpc, the non-linear effects on the volume averaged correlation function should be small. The power law approximation is in very good agreement with early results from the 2dF Galaxy Redshift Survey two-point correlation function out to 20$h^{-1}$Mpc (P. Norberg, private communication). Once the 2dF Galaxy Redshift Survey is finished, there will be no need to make this approximation.

$\bar{\xi}^s_{QSO}$ is found from the {\it Hubble Volume} by finding the number of $DD, DR$ and $RR$ pairs that have separations in the range 0-20$h^{-1}$Mpc and forming the two-point correlation using the estimator of Hamilton. These values agree well with the values obtained from the 2QZ 10K Catalogue as the mock correlation function was biased to approximately match early results from the data at $\sim$10$h^{-1}$Mpc.

For each cosmology, the bias between the galaxies and the mass at $z$=0, $b_{g \rho}(0)$, can be found from
\begin{equation}
b_{g\rho}(0) = \frac{\Omega_{\rm m}^{0.6}(0)} {\beta_{\rm g}(0)} .
\end{equation}
The real space galaxy correlation function can be determined from the redshift space galaxy correlation function
\begin{equation}
\bar{\xi}^r_g = \frac{\bar{\xi}^s_g}{[1 + \frac {2}{3} \beta_{\rm g}(0) + \frac{1}{5} \beta^2_{\rm g}(0)]},
\end{equation}
where the superscripts $r$ and $s$ indicate real and redshift space respectively.
The real space mass correlation function at $z$=0 can now be found as
\begin{equation}
\bar{\xi}^r_{\rho}(0) = \frac{\bar{\xi}^r_g(0)}{b_{g\rho}^2(0)}.
\end{equation}
The real space mass correlation function evolves according to linear theory such that
\begin{equation}
\bar{\xi}^r_{\rho}(z) = \frac{\bar{\xi}^r_{\rho}(0)} {G(z)^2},
\label{eq:evo}
\end{equation}
where $G(z)$ is the growth factor, which depends on cosmology, found from the formula of \citeasnoun{CPT92}. When $\Omega_{\rm m}(0)$=1, $G(z) = (1 + z)$.

We now relate the correlation function of the mass at $z$ measured in real space to the amplitude of the QSO clustering at $z$, measured in redshift space as we wish to know $\beta_{QSO}(\bar{z})$ as a function of $\Omega_{\rm m}(0)$. 
First we calculate $\Omega_{\rm m}(\bar{z})$ using
\begin{equation}
\Omega_{\rm m}(z) = \frac {\Omega_{\rm m}(0) (1 + z)^3} {\Omega_{\rm m}(0) (1 + z)^3 + \Omega_{\Lambda}(0) }.
\end{equation}
This is valid for flat cosmologies only. $\beta_{QSO}(z)$ is then given by
\begin{equation}
\beta_{QSO}(z) = \frac { \Omega_{\rm m}(z)^{0.6}} {b_{q\rho}(z)},
\label{eq:betaq}
\end{equation}
but as yet $b_{q\rho}(z)$ is unknown. $b_{q\rho}(z)$ is defined as
\begin{equation}
b^2_{q\rho}(z) = \frac{\bar{\xi}^r_{QSO}(z)}{\bar{\xi}^r_{\rho}(z)}
\end{equation}
substituting the above equations into equation \ref{eq:betaq} gives
\begin{equation}
\beta^2_{QSO}(z) = \Omega_{\rm m}(z)^{1.2} \frac { \bar{\xi}^r_{\rho}(z) }{ \bar{\xi}^r_{QSO}(z)} .
\label{eq:betaq2}
\end{equation}
$\bar{\xi}^r_{QSO}(z)$ can be found via 
\begin{equation}
\bar{\xi}^r_{QSO}(z) = \frac{\bar{\xi}^s_{QSO}(z)}{(1 + \frac{2}{3} \beta_{QSO}(z) + \frac{1}{5} \beta^2_{QSO}(z)) }.
\label{eq:xiqrxiqs}
\end{equation}
Substituting this into equation \ref{eq:betaq2}, gives
\begin{equation}
\beta^2_{QSO}(z) = \Omega_{\rm m}(z)^{1.2} \frac { \bar{\xi}^r_{\rho}(z) }{ \bar{\xi}^s_{QSO}(z)} [1 + \frac{2}{3} \beta_{QSO}(z) + \frac{1}{5} \beta^2_{QSO}(z)],
\end{equation}
a quadratic in $\beta^2_{QSO}(z)$ which can easily be solved, allowing $\beta_{QSO}(z)$ for any $\Omega_{\rm m}(0)$ to be found. We substitute $z=1.4$ to find the value of $\beta_{QSO}(z)$ at the average redshift of the survey.

The errors on $\beta_{QSO}(z)$ are found in the standard way, i.e. by differentiating $\beta_{QSO}(z)$ with respect to all the variables that contribute to the error and summing the components of the error in quadrature. In this case there are errors on $\beta_g(0)$, $\bar{\xi}^s_{QSO}(z)$ and $\bar{\xi}^s_g(0)$. The error on $\beta_g(0)$ is $\pm$ 0.07, the error on $\bar{\xi}_{QSO}(z)$ is the Poisson error found from the total $DD$ counts. The error on $\bar{\xi}_g$ is the least well known as the power law approximation is used. We assume an error of around 20\%, which is larger than the error expected from the final 2dF Galaxy Redshift Survey.

\begin{figure} 
\begin{centering}
\begin{tabular}{c}
{\epsfxsize=8truecm \epsfysize=8truecm \epsfbox[10 10 550 560]{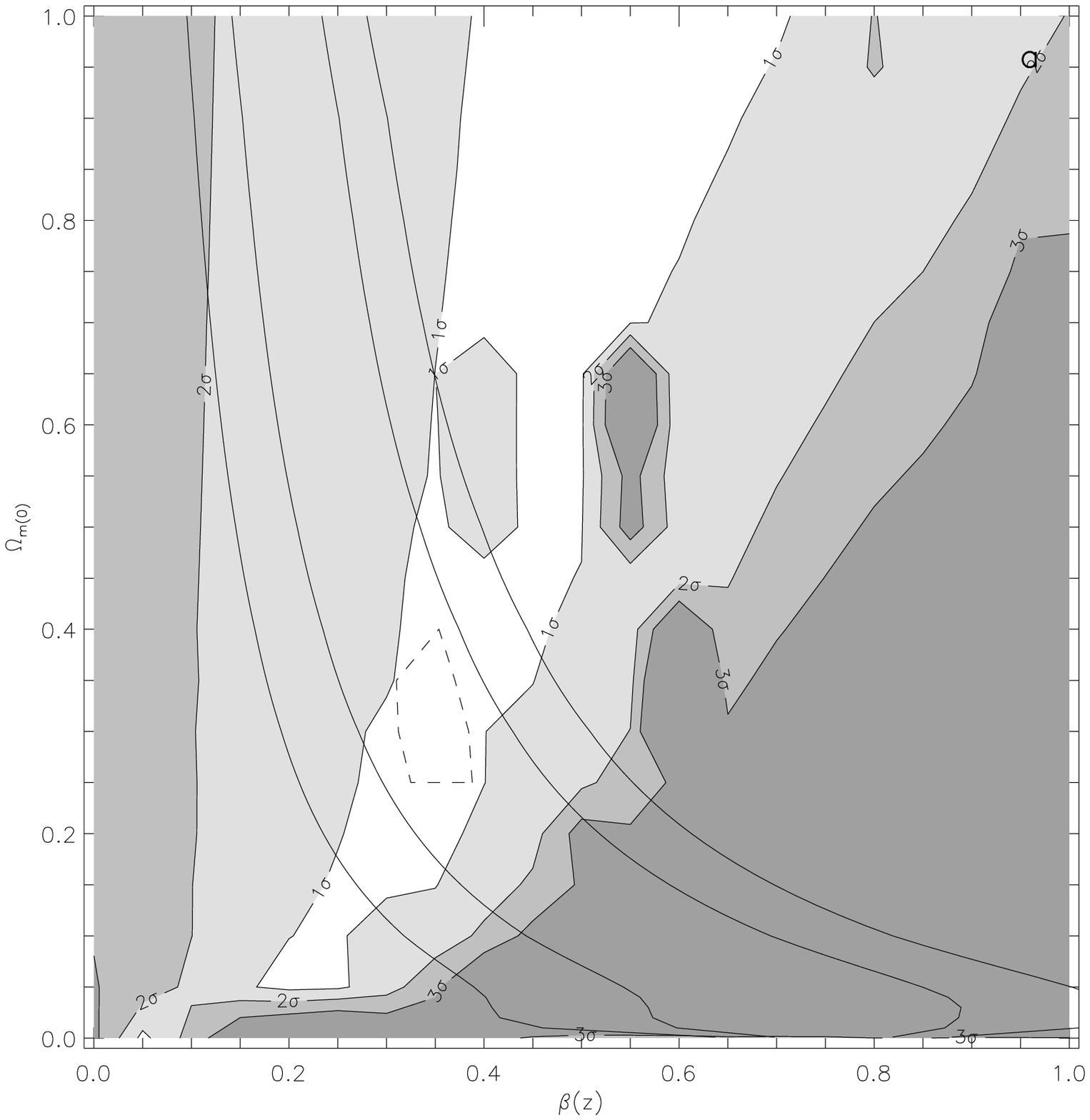}} \\
{\epsfxsize=8truecm \epsfysize=8truecm \epsfbox[10 10 550 560]{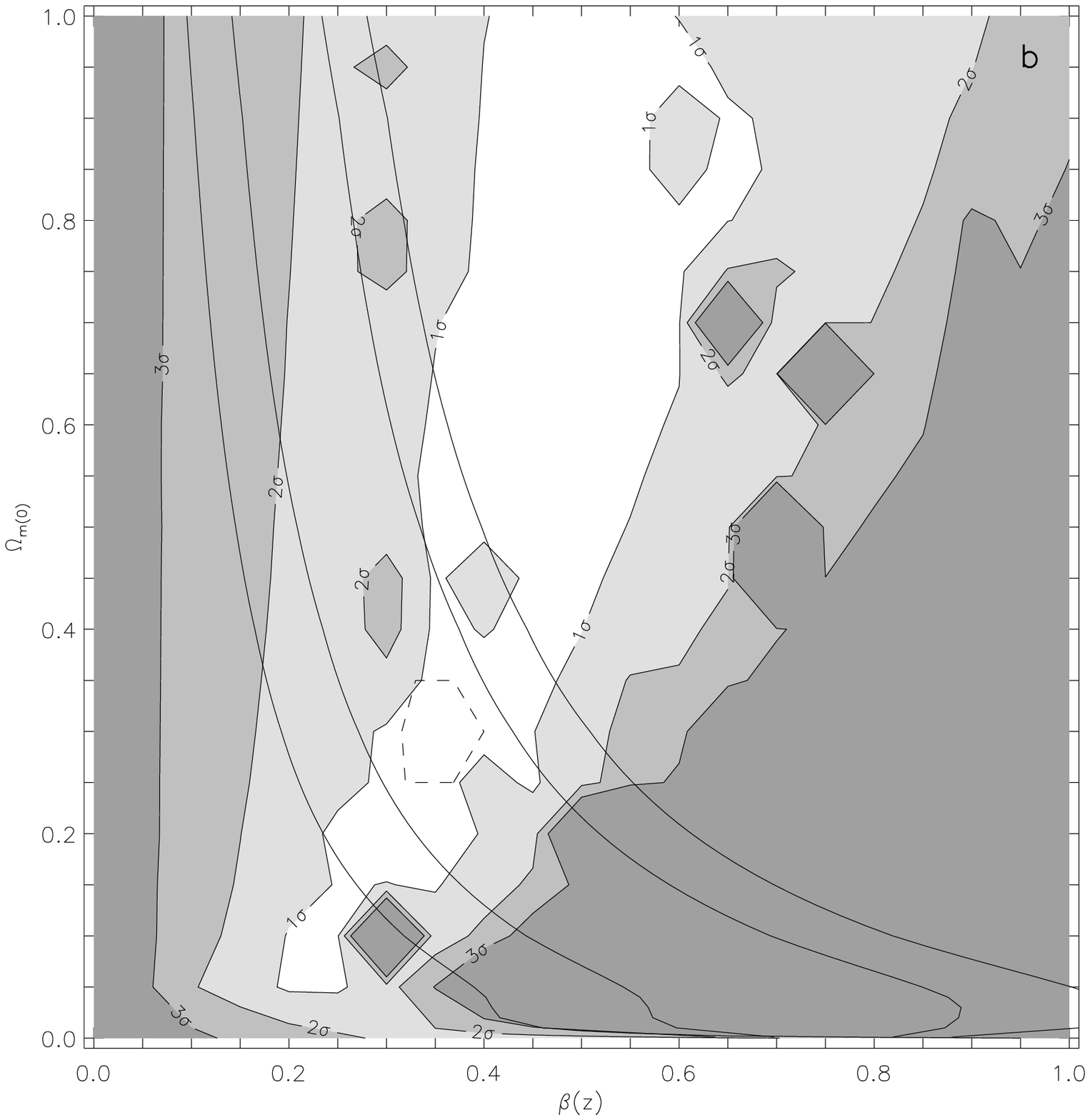}} \\
\end{tabular}
\caption
{Combining predicted results from measuring $\xi(\sigma,\pi)$ from the Hubble Volume mock catalogues of the full 2QZ Survey (contour plot as in Figure \ref{fig:HVres}) with results from the evolution of clustering. Plot (a) shows the contours with EdS assumed and plot (b) shows the contours with the $\Lambda$ cosmology assumed. The  lines show the 1 and 2$\sigma$ contours predicted from the method assuming $\beta_{\rm gal}$. The dashed lines show the 1$\sigma$ joint contours. }
\label{fig:comb}
\end{centering}
\end{figure}

\begin{figure} 
\begin{centering}
\begin{tabular}{c}
{\epsfxsize=8truecm \epsfysize=8truecm \epsfbox[10 10 550 560]{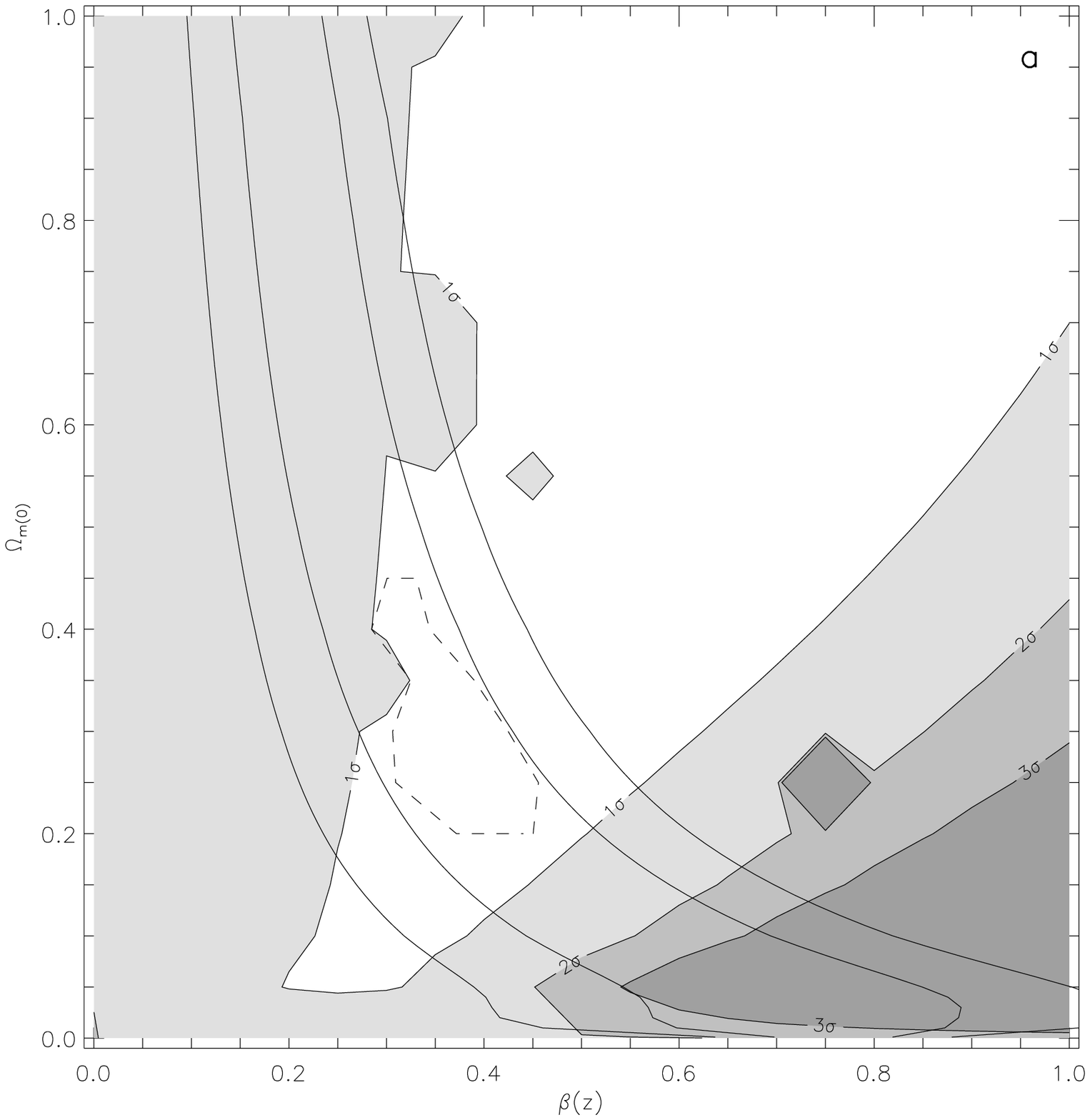}} \\
{\epsfxsize=8truecm \epsfysize=8truecm \epsfbox[10 10 550 560]{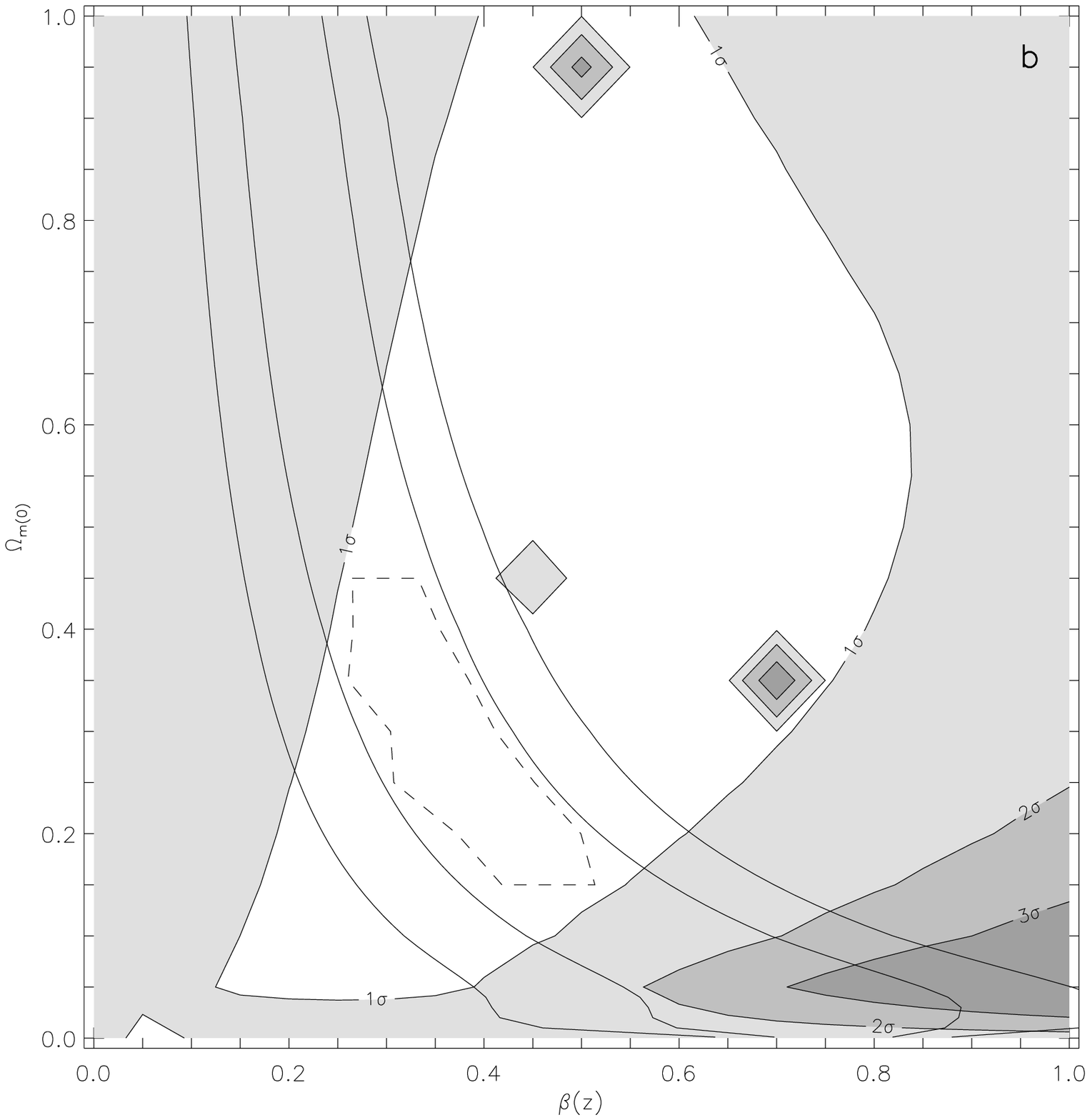}} \\
\end{tabular}
\caption{Figure (a) shows the constraints possible when we apply the 10k masks to the {\it Hubble Volume} cosmology. We assume the $\Lambda$ cosmology. The patchy angular selection function does not seem to bias the results. In (b) we apply the same analysis to the 10k masked {\it Hubble Volume} but we have added a 5$h^{-1}$Mpc Gaussian dispersion to the positions of the QSOs to mimic the redshift errors present in the 10k 2QZ sample. The dashed lines show the joint 1$\sigma$ contours. In both cases we are still able to recover the cosmology of the {\it Hubble Volume}. }
\label{fig:10ktest}
\end{centering}
\end{figure}

\subsection{Model Results}

A comparison between the constraint on $\Omega_{\rm m}(0)$ and $\beta_{QSO}(\bar{z})$ found here and the constraint found from fitting models to the full 2QZ {\it Hubble Volume} mock catalogue $\xi(\sigma,\pi)$ is shown in Figure \ref{fig:comb}. The contours are shown as in Figure \ref{fig:HVres} and the solid lines show the 1$\sigma$ and 2$\sigma$ constraints from the above method. We show the comparison for the two assumed cosmologies. It can be seen that $\Omega_{\rm m}(0)$=0, $\Omega_{\Lambda}(0)$=1 and $\Omega_{\rm m}(0)$=1, $\Omega_{\Lambda}(0)$=0 are ruled out with $\sim 2 \sigma$ confidence when the results are combined. The value of $\Omega_{\rm m}(0)$ is now constrained to be $\sim$0.3$^{+0.20}_{-0.15}$, by considering where the 1$\sigma$ contours from the two estimates overlap.

The constraints on $\beta_{QSO}(\bar{z})$ are even stronger than the constraints on cosmology. $\beta_{QSO}(\bar{z})$ can be constrained from the $\xi(\sigma,\pi)$ contours alone but by combining the errors, $\beta_{QSO}(\bar{z})$ can be measured to an accuracy of $\sim \pm$0.1 which places joint constraints on cosmology and the QSO-mass bias at the very least. This suggests that when 2QZ is finished, a powerful method for constraining cosmological parameters will be available.

\section{The 10k Catalogue}
\label{sec:10k}

Our overall aim here is to apply the analysis described above to the 10k catalogue of 2QZ QSOs \cite{Scott10k}. The catalogue contains 10681 QSOs and is publically available at {\tt www.2dfquasar.org}. This catalogue contains the most spectroscopically 
complete fields (i.e. fields in which more than 85 per cent of objects have 
been identified) that were observed prior to November 2000.
 
As the survey is not finished, it currently has a patchy angular selection function. A 1' by 1' completeness map is constructed for both the north and south galactic cap (NGC and SGC hereafter) regions to allow us to construct a random catalogue with the same angular selection function. We also take into account the effects of galactic extinction in the construction of the random catalogue. Full details are given in \citeasnoun{Scott00} and \citeasnoun{HoyleQSOPK}. 

There is an uncertainty in the measurement of the QSO redshifts due to the resolution and signal to noise of the 2dF spectra. Redshifts can currently only be measured to an accuracy of at best $\sigma(z)=0.0035$ \cite{Scott10k}. This means that as well as the intrinsic small scale `Finger of God' velocity dispersion, there is also an apparent velocity dispersion of around 600 km s$^{-1}$. If QSOs have a similar velocity dispersion as galaxies of $\sim 500$ km s$^{-1}$ (Ratcliffe et al. 1998c) then the observed velocity dispersion would be around 700-800 km s$^{-1}$.

\begin{figure*} 
\begin{centering}
\begin{tabular}{cc}
{\epsfxsize=8truecm \epsfysize=8truecm \epsfbox[10 10 550 560]{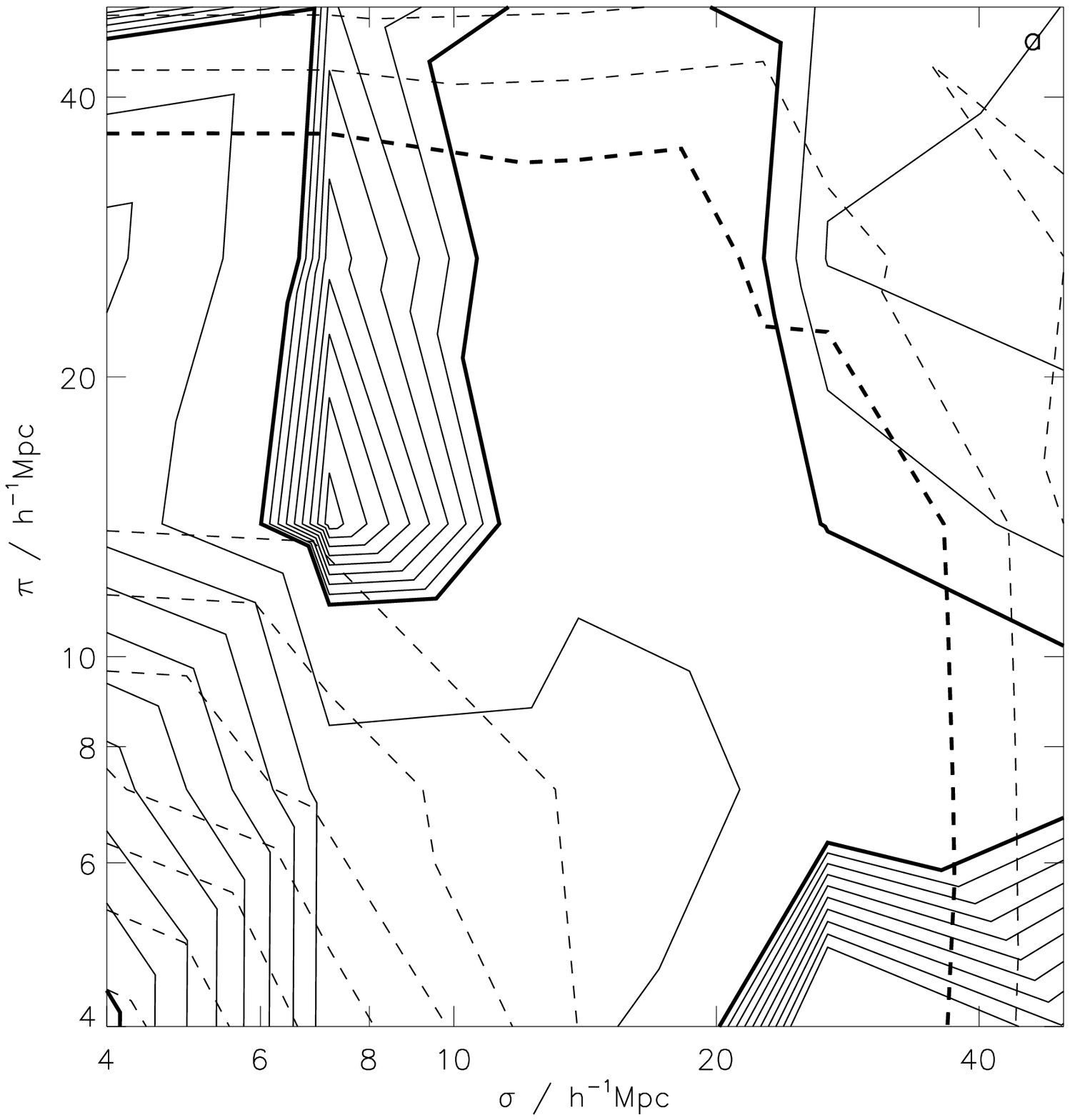 }} &
{\epsfxsize=8truecm \epsfysize=8truecm \epsfbox[10 10 550 560]{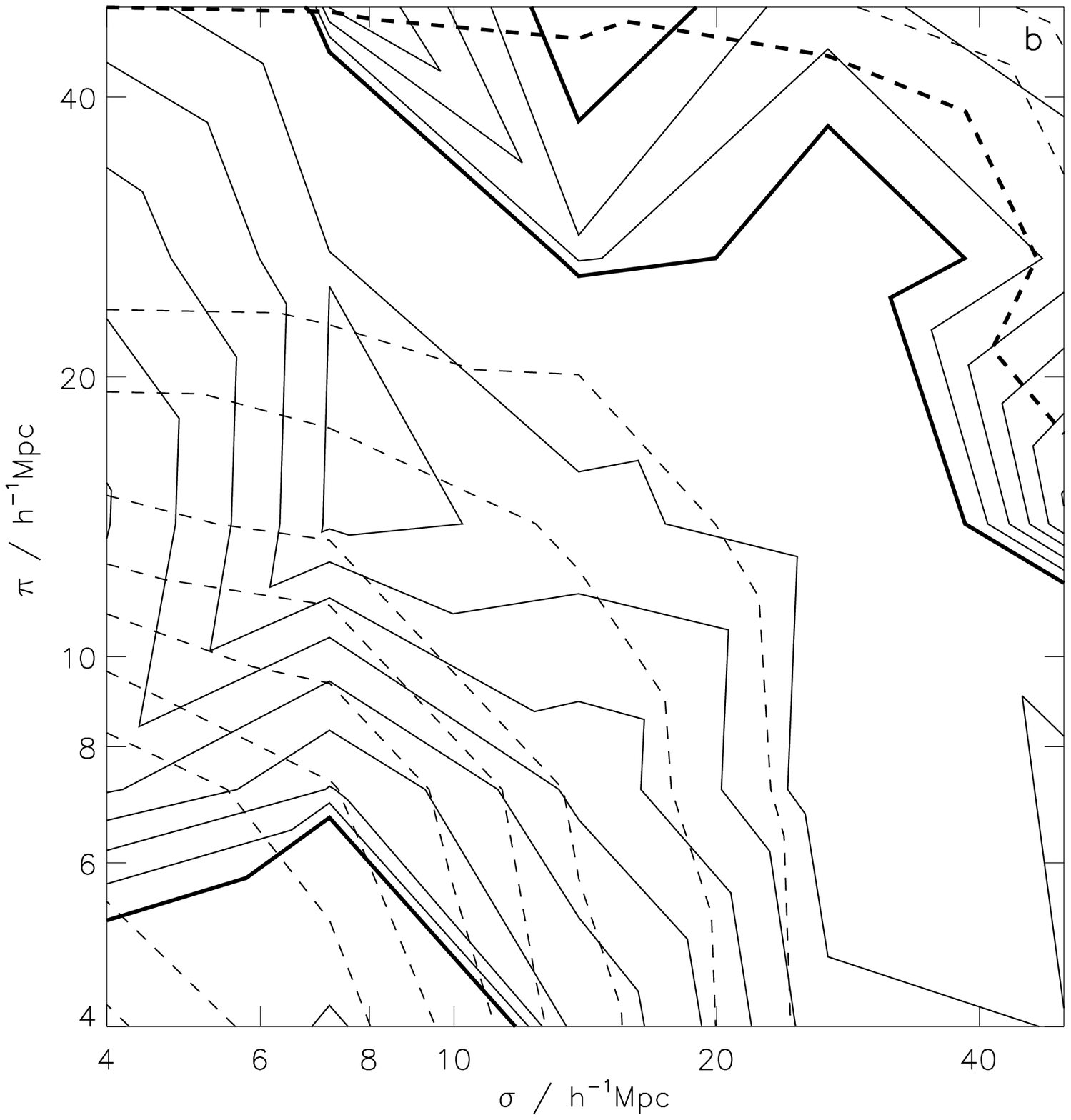}} \\
\end{tabular}
\caption{The solid lines show $\xi(\sigma,\pi)$ measured from the 10k data catalogue assuming the EdS cosmology (a) and the $\Lambda$ cosmology (b). The dashed lines show models. The models have the best values of $\Omega_{\rm m}(0)$ and $\beta_{\rm QSO}(\bar{z})$ found by combining the two techniques of constraining cosmology. When we assume EdS the values are $\Omega_{\rm m}(0)$=0.20, $\beta_{\rm QSO}$=0.35, when we assume the $\Lambda$ cosmology the values are $\Omega_{\rm m}(0)$=0.3, $\beta_{\rm QSO}$=0.35. We set $<w^2_z>^{1/2}$=800 km s$^{-1}$ in the models. }
\label{fig:xisigpi10k}
\end{centering}
\end{figure*}

\subsection{10k Hubble Volume Mock Catalogue Analysis}
\label{sec:10kHVMC}

To test the effect of the current window function on the fitting of $\xi(\sigma,\pi)$, we have again used the {\it Hubble Volume}. We imprint each of the completeness maps (NGC and SGC) onto one of the {\it Hubble Volume} slices, measure $\xi(\sigma,\pi)$ and combine the results as we do with the data. Due to the smaller number of QSOs in the current catalogue, the measurement of $\xi(\sigma,\pi)$ is more noisy than that from the final catalogue will be. Therefore, rather than binning in $\delta {\rm log}(\sigma,\pi)=0.2$, as was done for the analysis of the 25k mock catalogues, we use slightly larger bins of $\delta {\rm log}(\sigma,\pi)=0.3$ for the 10k mock and 2QZ analysis. This increases the signal within each bin but reduces the number of bins available for comparison with the models. As the uncertainty in the redshifts increases the small scale peculiar velocities, we restrict our fitting range to 7$h^{-1}$Mpc $< \sigma, \pi < 40 h^{-1}$Mpc when considering the 10k mocks and 2QZ data to minimise the risk of the small scale peculiar velocities affecting our ability to detect the geometric distortions. 

We predict that the constraints possible from $\xi(\sigma,\pi)$ measured from the 10k 2QZ data will not be as strong as for the finished survey, however, when we combine the results with the results from the amplitude of QSO clustering, we still manage to recover the simulation cosmology, although the errors are slightly larger than for the prediction from the finished survey. The combined results give best fit values of $\Omega_{\rm m}(0)$=0.25$^{+0.20}_{-0.20}$ $\beta_{\rm QSO}(\bar{z})$=0.40$^{+0.10}_{-0.15}$, shown in Figure \ref{fig:10ktest}(a). We just show the results assuming the $\Lambda$ cosmology for this test in Figure \ref{fig:10ktest}.

We also test how much the redshift errors may be affecting the constraint on cosmology. We take the positions of the mock QSOs from the {\it Hubble Volume} with the current mask applied and add a 5$h^{-1}$Mpc Gaussian dispersion to the line-of-sight distances and repeat the analysis. The results are shown in Figure \ref{fig:10ktest}(b). The constraint is slightly weaker than when we just apply the window but is still consistent with the values from the {\it Hubble Volume}. The best fit values are $\Omega_{\rm m}(0)$=0.20$^{+0.25}_{-0.10}$ $\beta_{\rm QSO}(\bar{z})$=0.40$^{+0.20}_{-0.20}$. We model the peculiar velocities using an exponential model, which has been found to be a better fit to N-body simulations (e.g. Cole, Fisher \& Weinberg 1994). However, if the velocities are dominated by redshift errors, a Gaussian model may be more appropriate. The fact that we are able to recover the {\it Hubble Volume} parameters using the exponential model even though we have added a Gaussian component to the small scale peculiar velocities suggests that the choice of model for the small scale peculiar velocities is not crucial for placing constraints on cosmology and $\beta_{\rm QSO}(z)$.

\subsection{Results from the 10k Catalogue}
\label{sec:10kres}

\begin{figure} 
\begin{centering}
\begin{tabular}{c}
{\epsfxsize=8truecm \epsfysize=8truecm \epsfbox[10 10 550 560]{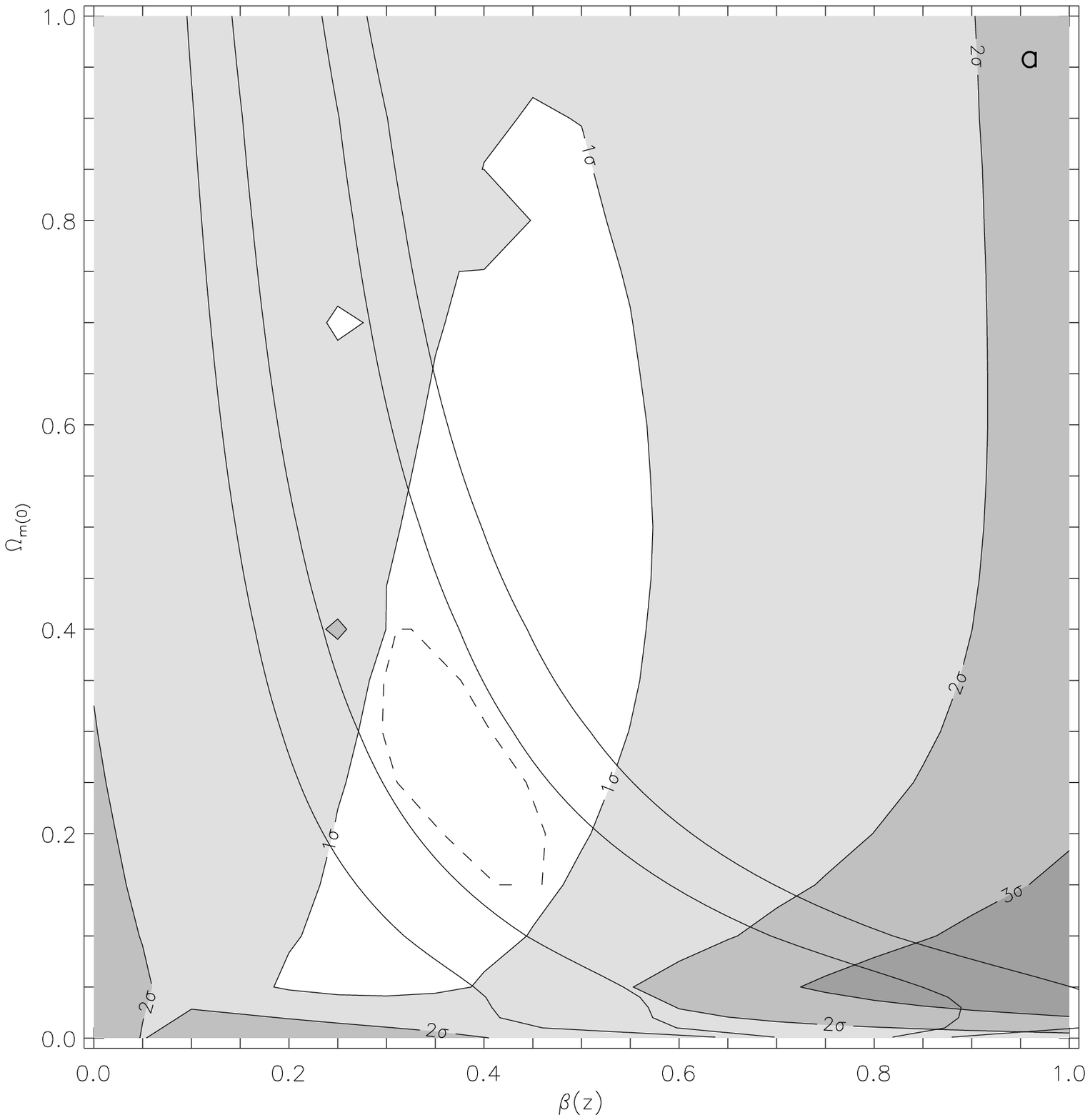}} \\
{\epsfxsize=8truecm \epsfysize=8truecm \epsfbox[10 10 550 560]{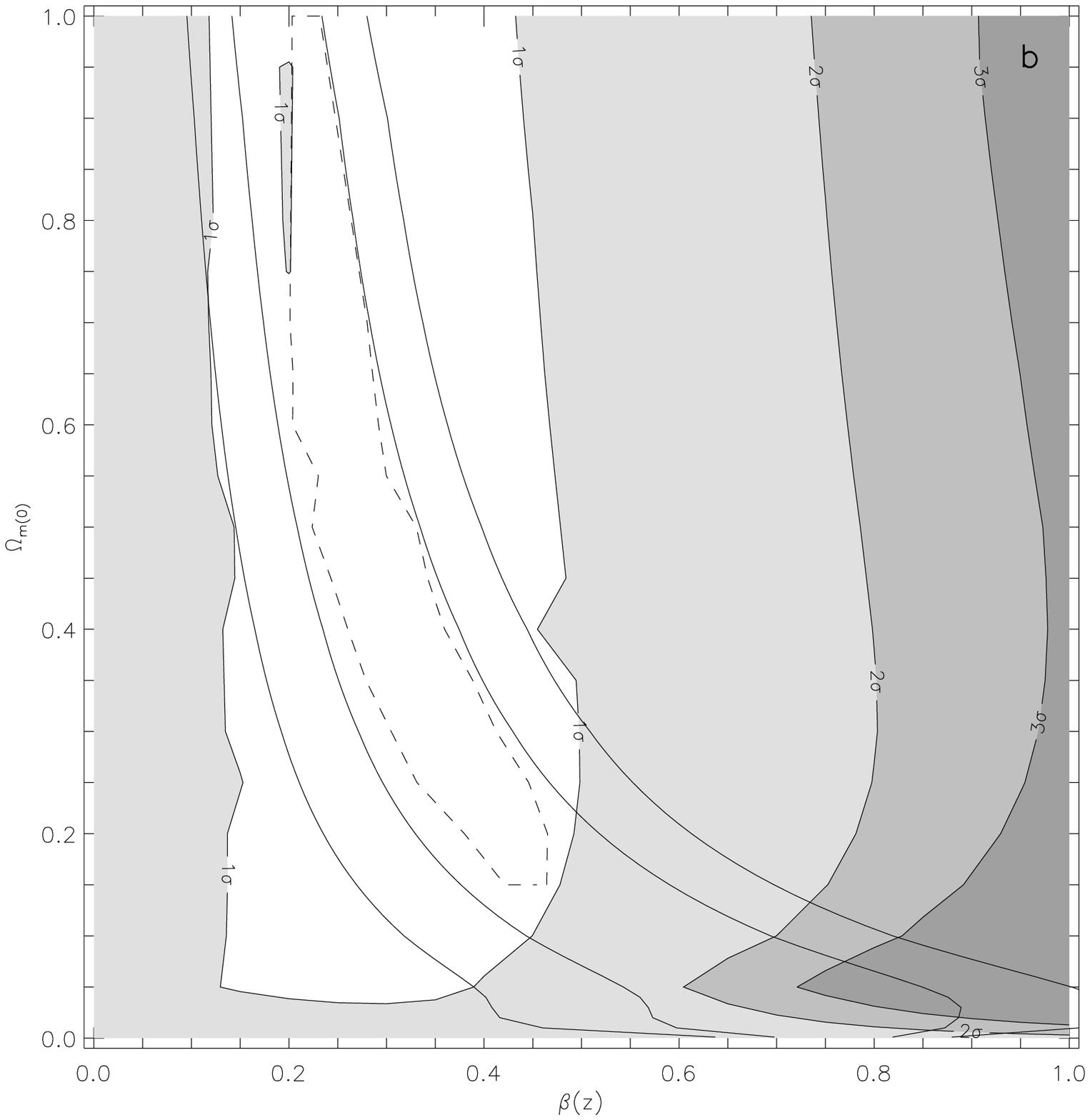}} \\
\end{tabular}
\caption
{Constraints on cosmology from $\xi(\sigma,\pi)$ measured from the 2QZ 10k data (greyscale) and from the evolution of the QSO clustering amplitude (solid lines) assuming the two different cosmologies, EdS (a) and $\Lambda$ (b). The dashed lines show the joint 1$\sigma$ contours.}
\label{fig:10kres}
\end{centering}
\end{figure}

We show $\xi(\sigma,\pi)$ measured from the 10k catalogue in Figure \ref{fig:xisigpi10k}, shown by the solid lines. The dashed lines show models, discussed below. There is evidence in the $\xi(\sigma, \pi)$ plots that the velocity dispersion is indeed higher than 400 km s$^{-1}$ as the data appears more elongated on small $\sigma$ scales than $\xi(\sigma, \pi)$ from the {\it Hubble Volume}. We have therefore considered models with $<w^2_z>^{1/2}$ of 400, 800 and 1200 km s$^{-1}$. There is very little difference between the values of $\chi^2$ found using each of the values. We obtain the best fit with $<w^2_z>^{1/2}$=800 km s$^{-1}$ but the constraints on cosmology are very similar for each value, especially when the results are combined with the evolution of clustering constraints. It is possible that the redshift errors will be reduced by the time the final sample is complete but for now, we only fit the models to the data on scales larger than 7$h^{-1}$Mpc to limit the effect of the small scale peculiar velocities on the model. The models in  Figure \ref{fig:xisigpi10k} and the fits shown in Figure \ref{fig:10kres} are done assuming $<w^2_z>^{1/2}$=800 km s$^{-1}$. 

The favoured model, if the $\Lambda$ cosmology is assumed, is $\Omega_{\rm m}(0)$=0.3 $\beta_{\rm QSO}(\bar{z})$=0.35. If the EdS cosmology is assumed, then $\Omega_{\rm m}(0)$=0.35 $\beta_{\rm QSO}$=0.45 is the favoured model. However, as can be seen from Figure \ref{fig:10kres}, the results are highly degenerate and a wide range of models are allowed. We do, however, constrain $\beta_{\rm QSO}(\bar{z})$ to be in the range 0.1$<\beta_{\rm QSO}<0.5$ at the $1 \sigma$ level. Currently, the only constraint we can place on cosmology from fitting $\xi(\sigma,\pi)$ alone is that $\Omega_{\rm m}(0)=0$ is marginally rejected. All other values of  $\Omega_{\rm m}(0)$ are contained in the 1$\sigma$ contour. 

If we combine the results with the evolution of clustering (Figure \ref{fig:10kres}) we find the results marginally favour low values of $\Omega_{\rm m}(0)$. The dashed lines show the joint 1$\sigma$ contours, found by adding the two values of the standard deviation together in quadrature. If the $\Lambda$ cosmology is assumed then we obtain a value of $\Omega_{\rm m}(0)$=0.3$^{+0.60}_{-0.15}$, $\beta_{\rm QSO}$=0.35$^{+0.10}_{-0.15}$. In this case $\Omega_{\rm m}(0)$=1 cannot be ruled out. If we assume EdS then slightly lower values of $\Omega_{\rm m}(0)$=0.20$^{+0.20}_{-0.10}$, $\beta_{\rm QSO}$=0.35$^{+0.10}_{-0.10}$ are favoured. These two models are shown by the dashed lines in Figure \ref{fig:10kres} (a, b).

Outram et al. (2001) have carried out a similar analysis of the 10k 2QZ data. They have analysed the power spectrum  measured parallel and perpendicular to the line of sight and they find that the best fitting results are $\Omega_{\rm m}(0)$=0.23$^{+0.44}_{-0.13}$ and $\beta_{\rm QSO}$=0.39$^{+0.18}_{-0.17}$.  Their results are very similar to those found here.

\section{Conclusions}
\label{sec:cnc}

We have developed a model for $\xi(\sigma,\pi)$ that takes into account the different types of distortions that are introduced into the clustering pattern from a redshift survey, such as small scale peculiar velocities, bulk motions and distortions due to the assumed cosmology differing from the underlying cosmology. We have checked that the analytic models are consistent with $\xi(\sigma,\pi)$ measured from the mock {\it Hubble Volume} catalogues and find that the model fits the simulation well over the range of scales $4 \lsim \pi, \sigma \lsim 40 h^{-1}$Mpc. 

We fit the models to $\xi(\sigma,\pi)$ measured from the {\it Hubble Volume} and confirm previous results that cosmology, $\Omega_{\rm m}(0)$ and $\Omega_{\Lambda}(0)$, will not be strongly constrained from the 2QZ Survey using geometric distortions in $\xi(\sigma,\pi)$. The errors are such that we cannot differentiate between the effects of $\Lambda$ and $\beta_{QSO}(\bar{z})$. 

Although cosmology cannot be directly obtained from the measurement of $\xi(\sigma,\pi)$ from the 2QZ Survey, constraints on $\beta_{QSO}(\bar{z})$ are possible from fitting models of $\xi(\sigma,\pi)$ to measurements of $\xi(\sigma,\pi)$ from the mock catalogues. This provides a joint constraint on the cosmology and the QSO-mass bias.

When these redshift space distortion results from $\xi(\sigma,\pi)$ are combined with information on the QSO-mass bias at $z=1.4$ from clustering evolution, new and direct constraints on cosmology  become available. The combined constraint immediately rules out values of $\Omega_{\rm m}(0)=$0 and $\Omega_{\rm m}(0)$=1 with 2$\sigma$ confidence, if we start with an $\Omega_{\rm m}(0)$=0.3, $\Omega_{\Lambda}(0)$=0.7 simulation.

We apply the methods to the 10k 2QZ catalogue and obtain encouraging results, although with only 10,000 QSOs currently any constraints on cosmology are limited. This analysis is consistent with flat models with 0.2$<\Omega_{\rm m}(0)<$1 and rejects flat models with $\Omega_{\Lambda}(0)$=1.

\section*{Acknowledgments}
We would like to recognise the considerable efforts of all the AAT staff responsible for the operation of the 2dF instrument which has made this survey possible. We thank our colleagues on the 2dF galaxy redshift survey team for facilitating many of the survey observations. We also thank the Virgo Consortium, especially Adrian Jenkins and Gus Evrard, for providing the {\it Hubble Volume} simulations.

\end{document}